\documentclass[journal=acsphotonics,manuscript=article,secnumarabic]{achemso}
\usepackage{chemformula} 
\usepackage[T1]{fontenc} 

\usepackage[utf8]{inputenc}
\usepackage{amsmath}
\usepackage{amsfonts}
\usepackage{amssymb}
\usepackage{array}

\usepackage{graphicx}
\usepackage{dcolumn}
\usepackage{bm}
\usepackage{color}
\usepackage{soul}
\usepackage{xcolor}
\usepackage{xr}
\usepackage{makeidx}
\usepackage{notes2bib}

\usepackage{xcolor}
\usepackage{mathtools}

\usepackage{titlesec} 
\titleformat{\section}{\normalfont\large\bfseries}{\thesection}{1em}{}
\setkeys{acs}{super=true}   
\newcommand{\nc}{\newcommand}

\nc{\nn}{\nonumber}
\nc{\XYZ}{ }
\def\e{\mathcal{E}}

\def\beq{\begin{equation}}
\def\eeq{\end{equation}}
\def\beqa{\begin{eqnarray}}
\def\eeqa{\end{eqnarray}}

\usepackage[normalem]{ulem}

\newcommand{\onlinecite}[1]{Ref.~[\hspace{-1 ex} \nocite{#1}\citenum{#1}]}

\author{Imon Kalyan}
\affiliation[BGU]{School of Electrical and Computer Engineering, Ben-Gurion University of the Negev, Israel}

\author{Ieng Wai Un}
\affiliation[SCNU1]{Guangdong Basic Research Center of Excellence for Structure and Fundamental Interactions of Matter, School of Physics, South China Normal University, Guangzhou 510006, China}
\alsoaffiliation[SCNU2]{Guangdong-Hong Kong Joint Laboratory of Quantum Matter, South China Normal University, Guangzhou 510006, China}

\author{Gilles Rosolen}
\affiliation[University of Mons]{Micro and Nanophotonic Materials Group,Research Institute for Materials Science and Engineering,University of Mons, 7000 Mons, Belgium}

\author{Nir Shitrit}
\affiliation[BGU]{School of Electrical and Computer Engineering, Ben-Gurion University of the Negev, Israel}

\author{Yonatan Sivan}
\affiliation[BGU]{School of Electrical and Computer Engineering, Ben-Gurion University of the Negev, Israel}
\email{sivanyon@bgu.ac.il}

\title{Photoluminescence from metal nanostructures - dependence on size}

\begin{document}

\maketitle

\begin{abstract}
For decades, there have been multiple seemingly contradicting experimental reports on the dependence of the photoluminescence from metal nanostructures on their size. We reconcile these reports using a simple analytic formula which is found to match well photoluminescence measurements for a range of structures and illumination conditions. Our expression requires only knowledge of the local electric field and temperature, and it points to the physical quantities that determine the emission strength and its dependence on size.
\end{abstract}

\section{Introduction}
The emission of light from metal nanostructures induced by illumination, frequently referred to colloquially as metal photoluminescence (PL), is a fundamental process occurring in a prototypical solid-state system\footnote{In many cases, the term photoluminescence is used to refer to spontaneous emission / radiative recombination of an electron and a hole, in contrast to inelastic light scattering, aka electronic Raman scattering. In that sense, in this work we use the term photoluminescence (PL) as a shorthand for emission, without distinguishing between these possibilities.}. It also has practical importance - metal nanoparticles are frequently used as non-bleaching fluorescent labels in bio-imaging{\XYZ ~\cite{metal_NP_PL_bioimaging,metal_NP_PL_bioimaging2,metal_NP_PL_bioimaging3}}, correlative light-electron microscopy~\cite{CLEM}, and more recently, as enablers of thermometry~\cite{Baumberg_SERS_T_measure,Cahill_T_measure,Orrit-Caldarola_T_measure,Kall_ACS_photonics_2018,Orrit-Caldarola_T_measure_transient}.

Despite that dual importance, it is surprising that many aspects of the metal PL have been under debate for decades. These debates are on, for example, whether the emission occurs due to recombination of an electron and hole residing both within the conduction band~\cite{Beversluis_PRB_2003,Lupton_transient_metal_PL_2017,Sivan-Dubi-PL_I} or involving a hole in the valence band (e.g.,~\onlinecite{Mooradian_metal_luminescence,Boyd-Shen-luminescence,Quidant_modes,Tigran-PL,Hecht_PRL-2019,Sheldon-nanoletters-2024}), on whether the emission statistics is fermionic or bosonic (e.g.,~\onlinecite{Orrit-Caldarola_T_measure,metal_luminescence_Link_ASE}), thermal or non-thermal (e.g.,~\onlinecite{metal_luminescence_Cahill_PNAS,Lupton_transient_metal_PL,Lupton_transient_metal_PL_2017,Nordlander_transient_e_dynamics,Sivan-Dubi-PL_I}), or how it depends on the local electric field (e.g.,~\onlinecite{metal_luminescence_Cahill_PNAS,Lupton_transient_metal_PL,Lupton_transient_metal_PL_2017,Giessen_ACS_photonics_2021}). Of particular interest is the dependence on the size of the nanostructure (e.g.,~\onlinecite{Feldman_QY_metal_PL,Orrit_QY_metal_PL_2011,Joel_Yang_ACS_Nano_2012,Tigran-PL,Hecht_PRL-2019}). Many early works studied the emission from metal clusters or few nm particles, primarily focussing on atomistic considerations and ligand coverage; these works highlighted the trade-off between the presence of a plasmonic resonance and the quantum yield for emission, see~\onlinecite{chinese_review_PL_few_nms}. For larger structures, there have been reports of contradicting trends. For example, Dulkeith {\em et al.}~\onlinecite{Feldman_QY_metal_PL} showed that the (time and) frequency integrated Stokes emission (SE) from spheres of growing radii illuminated by a short pulse scales linearly with their volume; similar results were reported by Gaiduk {\em et al.}~\onlinecite{Orrit_QY_metal_PL_2011} for continuous wave (CW) illumination. On the other hand, Lin {\em et al.}~\onlinecite{intraband_PL_Xiamen} and more recently, Bowman {\em et al.}~\onlinecite{Abajo_Tagliabue_PL} have demonstrated that the CW PL from rods and thin Au films, respectively, decreases with growing thickness; similar trends were reported in~\onlinecite{Hecht_PRL-2019,Abajo_DTU_PL} for pulsed illumination. Other works reported a mixed trend (e.g.,~\onlinecite{Joel_Yang_ACS_Nano_2012}).

These debates originate from the different conditions under which the PL measurements have been made (in terms of the number, density and shape of the nanostructure geometry, the illumination duration, intensity and wavelength, the relative position of the latter with respect to the resonances of the structure, the relative spectral position of emission with respect to the illumination wavelength, the geometry and thermal properties of the surrounding etc.) or the exact quantity studied {\XYZ (temporally and/or spectrally resolved/integrated PL)}, but also from the absence of a simple theory that encompasses both photonic aspects of the problem (via the photonic density of states, as appearing in Purcell's formula, see e.g.,~\cite{Nordlander_transient_e_dynamics}), as well as the thermal aspects and the electronic aspects, i.e., the distribution and associated weights of the continuum of possible recombination transitions.

Recently, Dubi and Sivan employed a simple analytic expression for the steady-state electron distribution in a Drude metal to derive an equally simple expression for the PL from metals illuminated by CW light~\cite{Sivan-Dubi-PL_I}. By relying on the empiric values of the permittivity to determine the absorption, this approach circumvents the need to specify the transition matrix element and circumvents the arguments about the origins of the emission (being radiative recombination or electronic Raman / inelastic light scattering etc.~\cite{Baumberg_SERS_T_measure,hot_es_Atwater,Abajo_Tagliabue_PL,metal_luminescence_Link_2023,Baumberg-Aizpurua_Nat_comm_2024_inelastic,Koopman_Auger_PL_flowers}). This {\XYZ predicted behaviour was verified experimentally in~\onlinecite{Sheldon-nanoletters-2024}, and} enabled the resolution of many of the disagreements described above (in particular, associated with the emission statistics and electric-field dependence). This expression was more recently extended to explain the dependence of the PL following short pulse illumination on the electric field~\cite{Sivan_tPL}.

In this work, we use the approach of~\onlinecite{Sivan-Dubi-PL_I} to reconcile the various seemingly contradicting reports on the size-dependence of the PL from the prototypical geometries of metal particles and films. Specifically, in Section~\ref{theory}, we combine the analytic expression of~\onlinecite{Sivan-Dubi-PL_I} with the detailed analysis of the heating of illuminated metal spheres~\cite{Un-Sivan-size-thermal-effect} and most importantly, with recent progress made by Loirette-Pelous and Greffet who showed how to determine the total PL of a metal nanostructure in the case of a non-thermal distribution~\cite{Pelous-Greffet}.

In Section~\ref{sec:results}, we use the resulting analytic expression to compute the PL from metal nano-spheres and nano-films as a function of their size. We distinguish between three cases. For weak illumination (hence, negligible heating), the dependence of the PL spectrum (both the Stokes Emission (SE) and the anti-Stokes Emission (aSE)) on the structure size is determined by a single parameter - the absorption cross-section density (or equivalently, via Kirchhoff's Law, the emission cross-section density); {\XYZ the electronic contribution is size-independent in this regime}. Thus, generically, the PL scales with the {\em illuminated} volume, i.e., it is a volume effect for small sizes and it becomes a surface effect when the size exceeds the penetration depth of a few 10s of nm. However, in the latter case, the resonances characteristic of particles modify further the size-dependence of the PL and effectively dominate it. Indeed, a change of size causes the emission at a given frequency to shift in and then out of resonance.

For stronger illumination (hence, moderate heating), the size-dependence of the PL is amplified through the dependence of the Bose function on the (electron) temperature. This has a fairly small effect on the SE, but a large effect on the aSE due to the exponential dependence of the Bose function on the temperature, which itself scales with the absorption cross-section density. 
For even stronger heating, thermo-optic effects kick in, and cause the quality factor of the resonance to decrease~\cite{Sivan-Chu-high-T-nl-plasmonics,Gurwich-Sivan-CW-nlty-metal_NP}. This causes weaker (excessive) heating at resonance (away from resonance), and hence, has a complex effect on the PL, depending on the emission frequency (SE/aSE).

We then demonstrate a good {\em qualitative} match between the prediction of our model and experimental results for spheres and films; in fact, good {\em quantitative} matches are observed in most cases. We demonstrate different trends by looking also at other particle shapes (rods and nano disks), and find a good qualitative match, even in the presence of modest field and temperature non-uniformity levels. Remarkably, the agreement usually extends beyond the formal limits of the analysis, specifically, in the presence of interband emission events, which are not accounted for in our analytic expression.

In Section~\ref{sec:Discussion}, we discuss possible reasons for the occasional quantitative mismatches we observe, the implications of the results, their pros and cons compared with more sophisticated approaches in the literature, specifically, the rigorous momentum-space calculations provides in~\onlinecite{Abajo_DTU_PL,Abajo_Tagliabue_PL,Baumberg-Aizpurua_Nat_comm_2024_inelastic}, and mention possible extensions of our approach.

\section{Theory}\label{theory}
\subsection{A microscopic view}
{\XYZ Based on the CW solution for $f$ (the steady-state electron distribution) obtained in~\onlinecite{Dubi-Sivan,Dubi-Sivan-Faraday} for uniformly illuminated Drude metals and its experimental verification in~\onlinecite{Shalaev_Reddy_Science_2020,Dubi-Sivan-MJs,Gabelli,Kumagai-ACS-phot-2023}, in~\onlinecite{Sivan-Dubi-PL_I}, }Sivan and Dubi presented a quantitative theory for the PL from Drude metals under continuous wave (CW) illumination, showing that the {\em probability} of emitted photons per unit frequency is given approximately by
\begin{eqnarray} \label{eq:gamma_guess}
\Gamma^{em}({\bf r},\omega,\omega_L;\e_F) \cong \gamma_\e({\bf r},\omega, \omega_L;\e_F) \rho_e^2(\e_F) I_e(\omega,\omega_L,|{\bf E}_L(\omega_L,{\bf r})|^2), \label{Eq:Gamma1}
\end{eqnarray}
where
\begin{eqnarray}
I_e(\omega,\omega_L,|{\bf E}_L(\omega_L,{\bf r})|^2) = \int f(\e + \hbar\omega,|{\bf E}_L(\omega_L,{\bf r})|^2)[1 - f(\e,|{\bf E}_L(\omega_L,{\bf r})|^2)]d\e \label{Eq:I_e_def}
\end{eqnarray}
represents the electronic contribution to the emission formula and
\begin{eqnarray} \label{eq:gamma_e}
\gamma_\e ({\bf r},\omega,\omega_L,\e) = \frac{\pi \omega_L V^2}{\epsilon_0}| \vec{\mu}(\e, \e + \hbar \omega_L)|^2 \rho_{phot}({\bf r},\omega)
\end{eqnarray}
represents the emission rate of a single electron. In the above, ${\bf r}$ and $\omega$ are the emitter position vector and frequency, $\omega_L$ is the excitation frequency, $\e_F$ is the Fermi energy, $\rho_e$ is the electron density of states and ${\bf E}_L$ is the local field. $|\vec{\mu}(\e_f,\e_i)|$ is the transition dipole moment between electronic states with an initial energy $\e_i$ and final energy $\e_f$ (assumed to be energy-independent) and $\rho_{phot}({\bf r},\omega)$ is the local density of photonic states (LDOPS).

The electronic contribution of the emission formula (Eq.~(\ref{Eq:I_e_def})) was shown in~\onlinecite{Sivan-Dubi-PL_I} to consist of a series of Planck's black-body radiation-like terms $(\e_{BB})$, i.e.,
\begin{eqnarray} \label{eq:Ie_guess}
I_e(\omega,\omega_L,|{\bf E}_L(\omega_L,{\bf r})|^2) &\sim& I_e(\omega,\omega_L,T_e,|{\bf E}_L(\omega_L,{\bf r})|^2) \label{Eq:I_e} \\
&\sim& \langle \e_{BB}(\omega;T_e)\rangle + 2 \langle\e_{BB}(\omega - \omega_L;T_e)\rangle \delta_E + \langle \e_{BB}(\omega - 2\omega_L;T_e)\rangle \delta_E^2 + \cdots, \nn
\end{eqnarray}
where $\langle \e_{BB}(\omega,T_e) \rangle = \hbar \omega / \big(e^{\frac{\hbar \omega}{k_B T_e}} - 1\big)$ and
\begin{equation} \label{eq:delta_E}
\delta_E({\bf r};\omega_L,pol',\hat{k'}) = p_{abs}({\bf r};\omega_L,pol',\hat{k'})/p_{sat}.
\end{equation}
Here, {\XYZ $T_e$ is the (effective) electron temperature, extracted from the first energy moment of the electron distribution~\cite{Italians_hot_es,GdA_hot_es,Dubi-Sivan,Dubi-Sivan-Faraday}}; for CW illumination, the latter is nearly equal to the phonon temperature, however, since the PL literature is inconsistent on the matter, the notation adopted in this manuscript emphasizes that the dependence is on the electron temperature. Further, the absorbed power density ($p_{abs}$) is defined in terms of the local ${\bf E}_L({\bf r};\omega_L,pol',\hat{k'})$ as
\begin{equation}
p_{abs}({\bf r};\omega_L,pol,\hat{k'}) = \frac{\omega_L\epsilon_0}{2}\epsilon_m''(\omega_L)|{\bf E}_L({\bf r};\omega_L,pol',\hat{k'})|^2 =\alpha_{abs}({\bf r};\omega_L, pol',\hat{k'}) I_{in}({\bf r};\omega_L, pol',\hat{k'}), \label{eq_p_abs}
\end{equation}
where $\alpha_{abs}({\bf r};\omega_L,pol',\hat{k'})$ is the absorption cross-section density or, in more general terms, the absorbed power density per unit incident intensity $(I_{in})$ of polarization $pol'$ and incidence direction $\hat{k'}$ at a position ${\bf r}$ corresponding to the illumination frequency. The saturation power density can be approximated by~\onlinecite{Dubi-Sivan-Faraday,Sivan-Dubi-PL_I}
\begin{equation} \label{eq:p_sat}
p_{sat}(\omega_L) = \frac{3}{4}\frac{n_e(\hbar\omega_L)^2}{\e_F \tau_{e-e}},
\end{equation}
where $n_e$ is the electron density and $\tau_{e-e}$ is an average rate of collisions between electrons. For simplicity, we set $\tau_{e-e}$ in Eq.~(\ref{eq:p_sat}) to a value typical for a non-thermal electron, i.e., {\XYZ we use the Fermi Liquid theory expression~\cite{Quantum-Liquid-Coleman,Dubi-Sivan}} at $\e \sim \e_F + \hbar \omega$.

\begin{figure}[htp]
\centering
\includegraphics[width=8cm]{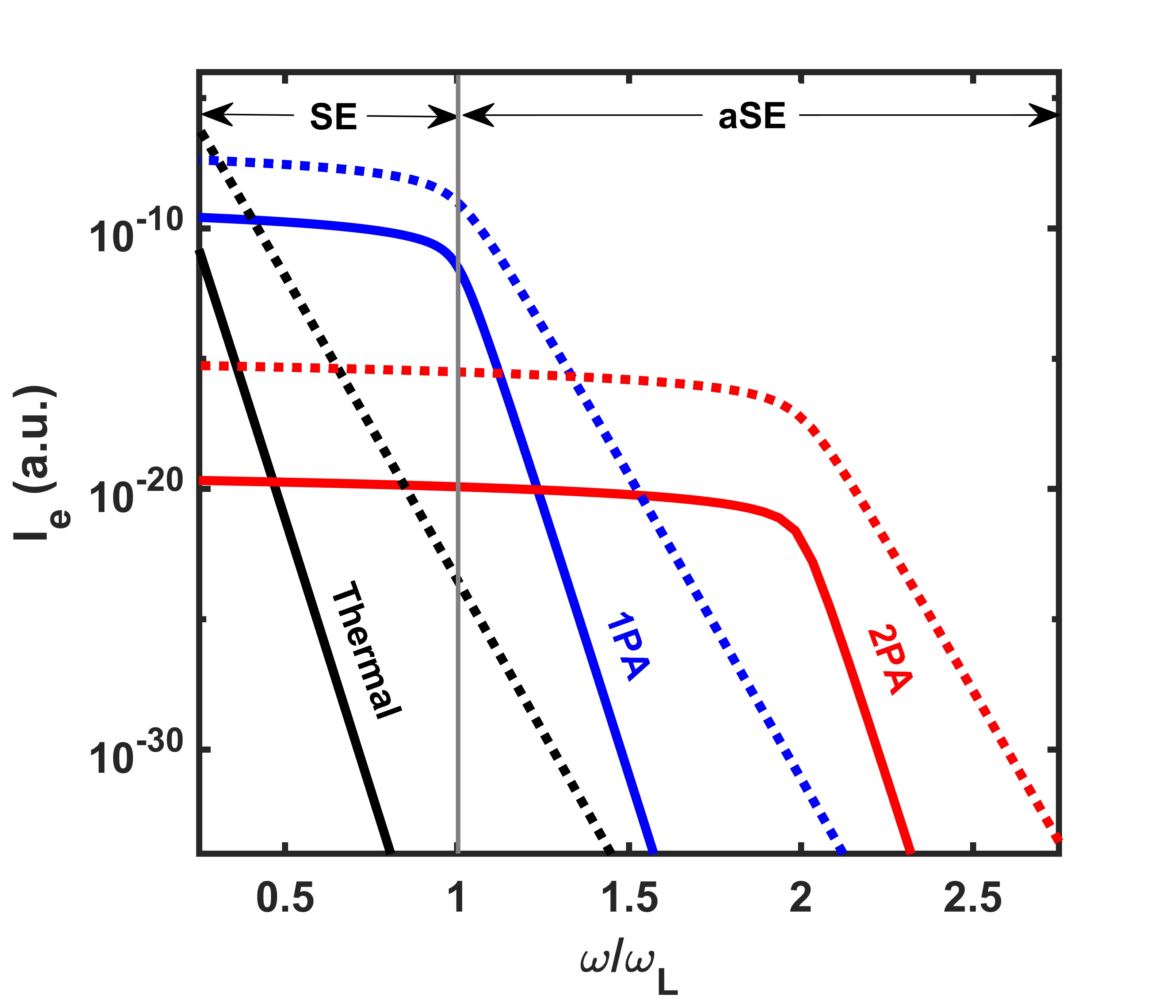}
\caption{ (Color online) Schematic of the contribution from thermal emission: $\langle \e_{BB}(\omega;T_e) \rangle $ (black), $1$PA: $2\langle \e_{BB}(\omega - \omega_L;T_e) \rangle \delta_E$ (blue), $2$PA: $\langle \e_{BB}(\omega - 2\omega_L;T_e) \rangle \delta_E^2$ (red) from Eq.~(\ref{Eq:I_e}) calculated under CW wave illumination. The plots are shown for two intensities, where solid and dotted lines indicate lower and higher intensities, respectively. } \label{Fig.Schematic_of_Ie}
\end{figure}

The first term on the RHS of Eq.~(\ref{eq:Ie_guess}) represents the average energy of thermal emission per electromagnetic mode (i.e., for vacuum electric fields); the next terms represent the non-thermal emission caused by deviations of the electron distribution from thermal equilibrium due to one photon absorption (1PA), two photon absorption (2PA) etc.. As discussed in~\onlinecite{Sivan-Dubi-PL_I,Sivan_tPL}, for CW illumination, these terms are typically small compared to the thermal emission at mid-IR frequencies, but dominate the emission close to the illumination frequencies and above them~\cite{Sivan-Dubi-PL_I,Sivan_tPL}.{\XYZ All these details are depicted in the schematic representation (Fig.~\ref{Fig.Schematic_of_Ie})}. The complete step structure of the non-thermal contributions was observed experimentally for CW illumination for the first time in~\onlinecite{Sheldon-nanoletters-2024}.

Notably, the emission~(\ref{eq:gamma_guess}) is assumed to occur only within the conduction band, i.e., to involve only intraband transitions; {\XYZ in that sense, strictly speaking, $\alpha_{abs}$ should include only the intraband contribution (as in~\onlinecite{Pelous-Greffet}). This would not matter for the illustrative examples below, for which we study Ag nanostructures, but could somewhat affect the match between our predictions and the experimental data (see Sections~\ref{subsub:exp_sphere} and~\ref{subsub:exp_layers}). In order to include interband absorption and emission transitions, our PL formula can be amended by accounting for the increased number of non-thermal holes above the interband absorption threshold in the electron distribution function $f$, as shown in Fig.~S10(b) of~\onlinecite{Abajo_Tagliabue_PL}, and adding the valence bands to the electron density of states}. As shown below, our analysis provides a qualitative (and usually, also a quantitative) match to the experimental data even without accounting for these additional transitions.

\subsection{A macroscopic view}
The result~(\ref{eq:gamma_guess}) applies for a general point within the metal. It was so far used to understand the general parametric dependence of the PL on the electric field, its statistics etc. with only a phenomenological address of the shape and size of the emitting nanostructure.

In order to account for a specific nanostructure geometry, one needs to sum over the random spontaneous emission events from the nanostructure volume as a function of space. This should allow one to account also for the {\em actual} magnitude of the emission, thus, including also the portion of the emitted photons that got re-absorbed (sometimes referred to as re-cycled), rather than only the {\em probability} of emission (as given by $\Gamma^{em}$~(\ref{eq:gamma_guess})). This can be done using the local Kirchhoff Law for non-isothermal bodies~\cite{Greffet_PRX_2018,abs_cross_sec_density_Greffet,abs_cross_sec_density2_Greffet} and its extension to metals having non-equilibrium electron distributions~\cite{Pelous-Greffet}. The latter study showed that the actual emitted power in the direction $\hat{k}$ into a solid angle $d\Omega$ ($ = \frac{d^3k}{k^2dk}$; see Eq.~(9.15.9) in~\onlinecite{Reif_Fund.stat._n_ther.phy.}) is given by
\begin{equation}
\begin{split}
dP^{em}(\omega,\omega_L, \hat{k}) \cong \frac{\rho_{phot}^{(0)}}{8\pi} \sum_{pol = s,p} \int_V \alpha_{abs}({\bf r};\omega,pol,-\hat{k})
{I_e(\omega,\omega_L,T_e({\bf r}),|{\bf E}_L({\bf r};\omega_L,pol',\hat{k'})|^2)}d^3r d\omega d\Omega, \label{Eq:P_em_main_old}
\end{split}
\end{equation}
where $\rho_{phot}^{(0)} = \frac{\omega^2}{\pi^2 c^2}$ is the photonic density of states of free space, $pol$ and $pol'$ are the polarization of the emitted and incident waves, respectively, and $\alpha_{abs}({\bf r};\omega,pol,-\hat{k})$ is the emission cross-section density (aka emissivity density), which by the local Kirchhoff Law, equals the absorption cross-section density. In that regard, following~\onlinecite{Greffet_PRX_2018,abs_cross_sec_density_Greffet,abs_cross_sec_density2_Greffet}, and unlike in~\onlinecite{{Pelous-Greffet}}, where the PL was expressed in terms of absorption cross-section, Eq.~(\ref{Eq:P_em_main_old}) utilizes the {\XYZ absorption cross-section density} to extend the formulation to structures with nonuniform field distributions.

Using Eqs.~(\ref{eq:Ie_guess})-(\ref{eq_p_abs}) in Eq.~(\ref{Eq:P_em_main_old}) gives
\begin{eqnarray}\label{Eq:P_em_main}
dP^{em}(\omega,\omega_L, \hat{k})
&\sim& \frac{\omega^2}{8\pi^3 c^2} \sum_{pol = s,p}~ \int_V \bigg[ {\frac{I_{in}}{p_{sat}}}\alpha_{abs}({\bf r};\omega,pol,-\hat{k}) \alpha_{abs}({\bf r};\omega_L,pol',\hat{k'}) \nn \\
&&2\langle\e_{BB}(\omega - \omega_L;T_e)\rangle_r + {\frac{I_{in}^2}{p^2_{sat}}}\alpha_{abs}({\bf r};\omega,pol,-\hat{k}) \alpha^2_{abs}({\bf r};\omega_L,pol',\hat{k'})\nn\\
&&\langle\e_{BB}(\omega - 2 \omega_L;T_e)\rangle_r+\cdots\bigg] d^3r d\omega d\Omega.
\end{eqnarray}

Eq.~(\ref{Eq:P_em_main}) shows that the quantities that determine the emission are the absorption cross-section density $\alpha_{abs}$ and the (electron) temperature distribution. {\XYZ This dependence is similar to that provided in~\onlinecite{Aizpurua_Ren_PL}, as well as to the scaling of surface-enhanced Raman Scattering (see~\onlinecite{Khurgin_Purcell}), but goes beyond these theories by providing a rigorous (rather than an empiric~\cite{Aizpurua_Ren_PL}) account of the effect of the electron distribution to the PL. } As shown below, the expression~(\ref{Eq:P_em_main}) also circumvents the more advanced yet complicated $k$-space calculations (as e.g., in~\onlinecite{Abajo_DTU_PL,Abajo_Tagliabue_PL,Baumberg-Aizpurua_Nat_comm_2024_inelastic}), without compromising much predictive capabilities.

We now note that the spatially-integrated emission~(\ref{Eq:P_em_main}) involves space-varying functions. Specifically, the spatial distribution of $\alpha_{abs}$ is determined by the material constituents and geometry of the nanostructure. In addition to gradients induced by illumination non-uniformity (e.g., when a focussed beam illuminates a film), Eq.~((\ref{Eq:P_em_main})) exhibits significant gradients on scales exceeding the penetration (skin) depth, i.e., for more than a few tens of nm for noble metals. In contrast, the non-uniformity of temperature is typically much weaker than that of the electric field (and hence $\alpha_{abs}$) due to the strong (electron, hence) heat diffusion in metals. This (electron) temperature non-uniformity should, in principle, be extracted from a self-consistent solution of microscopic equations for the electron dynamics; however, in practice, since the deviation from thermal equilibrium is minute for CW illumination~\cite{Dubi-Sivan,Dubi-Sivan-Faraday,Kumagai-ACS-phot-2023}, it is usually determined by coarse-graining such equations into heat equations, see, e.g.,~\onlinecite{delFatti_nonequilib_2000,GdA_hot_es,Dubi-Sivan,Dubi-Sivan-Faraday}. The resulting single or two temperature models reveal that heat gradients and electron-phonon temperature differences are very small (see, e.g.,~\onlinecite{thermo-plasmonics-review,ICFO_Sivan_metal_diffusion,Sivan_Spector_metal_diffusion,Gao_negative_diffusion_2023,Tielrooij_Sivan_negative_diffusion,Giri_Hopkins_diffusion}). Finally, like the absorption cross-section density $\alpha_{abs}$ and unlike the electron temperature $T_e$, the non-thermal part of the electron distribution, manifested via $\delta_E$~(\ref{eq:delta_E}), is roughly determined by the local electric field distribution. As predicted in~\onlinecite{hot_es_Atwater,Louie_DFT_Au_Ag,Sivan_non_thermal_e_transport} and demonstrated experimentally in~\onlinecite{Abajo_Tagliabue_PL}, the reason for that is the minimal (few nm) mean free path of these electrons, which in turn, originates from their femtosecond-scale collision rate. Accordingly, to determine the spatial dependence of the PL integrand, in what follows we use heat equations to determine the electron temperature and rely on the solution of Maxwell's equations to determine the electric field, and hence, $\alpha_{abs}$ and $\delta_E$. Having said that, in many cases (specifically, for few nm spheres and uniformly illuminated thin films, both studied below), $\alpha_{abs}$ is quite uniform. In such cases, $\alpha_{abs}$ can be estimated from measurable quantities like the absorption cross-section, $\sigma_{abs}$ (in the case of particles) or absorptance, $A$ (in the case of films), thus, simplifying the formulation slightly, and making the analysis of the PL simpler.

\section{Results}\label{sec:results}

\subsection{Nano-spheres under CW illumination}

\subsubsection{Determination of the electron temperature}\label{sec.T_sphere}
\begin{figure}[htp]
\centering
\includegraphics[width=16.5cm]{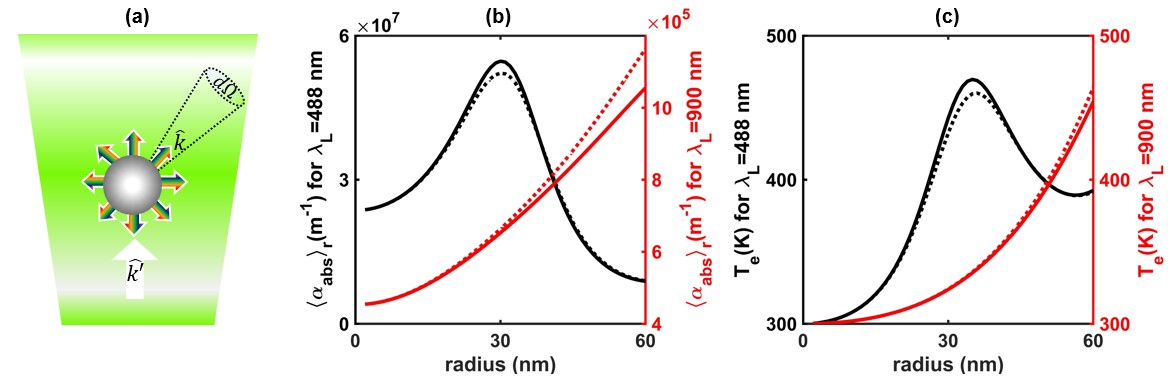}
\caption{ (Color online) (a) A schematic illustration of light emission from a sphere. (b) The absorption cross-section density $\langle \alpha_{abs}\rangle_r$ vs. radius for $\lambda_L = 488$ nm (black) and $900$ nm (red) for Ag spheres in oil. (b) The corresponding (electron) temperature reached due to CW illumination of the Ag spheres, calculated using Eq.~(\ref{Eq_T_sph_CW}). The corresponding illumination intensities are $I_{in} = 0.25$ MW/cm$^2$ (black line) and $I_{in} = 3.5$ MW/cm$^2$ (red line). The thermo-optic effect on $\langle \alpha_{abs}\rangle_r$ and $T_e$ is represented by the dotted lines. }
\label{Fig.alpha_and_T_sphere}
\end{figure}

We start our analysis by considering the PL from nano-spheres under CW plane wave illumination in a uniform optical and thermal environment (oil, in our case), see Fig.~\ref{Fig.alpha_and_T_sphere}(a). We choose silver as a prototypical plasmonic material, as it does not require accounting for interband transitions for illumination with visible light. As mentioned above, in this case, it is customary to neglect the small difference between the electron and phonon temperatures (see, e.g.,~\onlinecite{Abajo_nano-oven,Dubi-Sivan,Dubi-Sivan-Faraday}) and consider a single temperature model. Nevertheless, as mentioned, we choose to denote the temperature below as $T_e$, in order to emphasize that it is the electron temperature that is the relevant quantity as far as photon emission in the visible and near infrared spectral regimes is concerned. Then, the steady-state temperature $T_e({\bf r})$ can be obtained by solving the heat diffusion equation~\cite{thermo-plasmonics-basics}
\begin{equation}\label{Eq_T_Steady}
\begin{split}
\nabla \cdot [\kappa({\bf r})\nabla T_e({\bf r},\omega_L)] = \Biggl\{
\begin{matrix}
- \alpha_{abs}({\bf r};\omega_L) I_{in},\quad \text{inside the nanostructure,} \\
0, \quad\quad\quad\quad\quad\quad
\text{outside the nano structure,}
\end{matrix}
\end{split}
\end{equation}
where $\kappa$ is the thermal conductivity and we used Eqs.~(\ref{eq_p_abs}) on the right-hand-side. Due to the symmetry, $\alpha_{abs}({\bf r};\omega_L)$ for a nanosphere is independent of the direction and polarization of the incident beam.

As shown already in~\onlinecite{thermo-plasmonics-basics,Un-Sivan-size-thermal-effect}, even in the presence of substantial field non-uniformity (occurring for nano-spheres of more than a few tens of nm in size), the high thermal conductivity of the metal ensures that the temperature of the nanosphere is only weakly inhomogeneous. In fact, as shown in~\onlinecite{Un-Sivan-size-thermal-effect}, the temperature can be calculated quite accurately also when replacing $\alpha_{abs}({\bf r};\omega_L)$ in Eq.~(\ref{Eq_T_Steady}) by its spatial average, $\langle \alpha_{abs}({\bf r};\omega_L)\rangle_r = \frac{1}{V} \int \alpha_{abs}({\bf r};\omega_L) d{\bf r} = \sigma_{abs}(\omega_L)/V$. Particularly, for spheres whose radius is up to $\approx 60$ nm, the temperature can be approximated as
\begin{equation}\label{Eq_T_sph_CW}
T_e \approx \langle T_e\rangle_r = T_h + \frac{\langle \alpha_{abs}({\bf r};\omega_L)\rangle_r I_{in}}{3 \kappa_h} a^2.
\end{equation}

Here, $\kappa_h$ and $T_h$ are the thermal conductivity and the temperature of the surrounding media far from the particle.

As a representative example, Fig.~\ref{Fig.alpha_and_T_sphere}(b)-(c) show $\langle \alpha_{abs}({\bf r};\omega_L)\rangle_r$ and the (electron) temperature~(\ref{Eq_T_sph_CW}), respectively, calculated for Ag spheres of sizes up to $60$ nm dispersed in oil (permittivity $\epsilon_h = 2.235$, thermal conductivity $k_h = 0.2873$ Wm$^{-1}$K$^{-1}$) for an excitation laser wavelength $\lambda_L$ close to the (dipolar) plasmon resonance ($488$ nm) and for a wavelength farther from the resonance ($900$ nm). The absorption cross-section (density) is well studied~\cite{Bohren-Huffman-book}, hence, described below only briefly. Specifically, $\langle \alpha_{abs}({\bf r};\lambda_L = 488 nm)\rangle_r$ shows a peak around $30$ nm. For $a < 10$ nm, the absorption cross-section $\sigma_{abs}(\omega_L)$ (hence, $\langle \alpha_{abs}({\bf r};\omega_L)\rangle_r \sim \sigma_{abs}(\omega_L) / a^3$) takes the quasi-static form, $\sigma_{abs}^{qs} \simeq \frac{6\pi}{\lambda} V {\Im{\left(\frac{\epsilon - \epsilon_{bg}}{\epsilon + 2\epsilon_{bg}}\right)}}$ (see Fig.~\ref{Fig.ACS_vs_radius} or more generally,~\onlinecite{Bohren-Huffman-book}). As the sphere radius increases, the $\lambda \sim 400$ nm dipole resonance undergoes a red-shift (see Fig.~\ref{Fig.alpha_vs_WL}(a)). Thus, since the chosen excitation wavelength $\lambda_L \sim 488$ nm is at the long wavelength tail of the resonance, $\langle \alpha_{abs}({\bf r};\omega_L)\rangle_r$ increases with $a$ as the dipole resonance tunes towards $\lambda_L$. However, the strength of the dipole resonance decreases with growing sphere size, so that beyond $a \sim 30$ nm, the importance of higher-order resonances increases, and thus, the absorption becomes limited to the surface. The corresponding size dependence of $\sigma_{abs}$ is shown in Fig.~\ref{Fig.ACS_vs_radius}.

Similarly, for excitation at $\lambda_L = 900$ nm, the averaged absorption cross-section density $\langle \alpha_{abs}\rangle_r$ increases monotonically with the sphere size for all sizes studied here since the peak response occurs beyond the range of consideration in this work, see Fig.~\ref{Fig.ACS_vs_radius}(b). In that sense, the off-resonance illumination case behaves as in the small size regime of the on-resonance case; we will not dwell on it further in this study.

Fig.~\ref{Fig.alpha_and_T_sphere}(c) shows the corresponding size-dependence of the electron temperature. By Eq.~(\ref{Eq_T_sph_CW}), and as explained in~\onlinecite{Un-Sivan-size-thermal-effect}, the size-dependence of the temperature $T_e$ originates from the proportionality with $\langle \alpha_{abs}({\bf r};\omega_L)\rangle_r$ and $a^2$. Specifically, for on-resonance excitation, the product $\langle \alpha_{abs}({\bf r};\omega_L)\rangle_r a^2$ results in a peak of $T_e$ at $\approx 35$ nm while for off-resonance excitation, the monotonic increase of $\langle \alpha_{abs}({\bf r};\omega_L)\rangle_r$ results in a monotonic increase of $T_e$. The similarity to the size-dependence of the absorption cross-section density is apparent.

\subsubsection{Determination of the PL}\label{sec_PL_sphere}
\begin{figure}[htp]
\centering
\includegraphics[width=16.5cm]{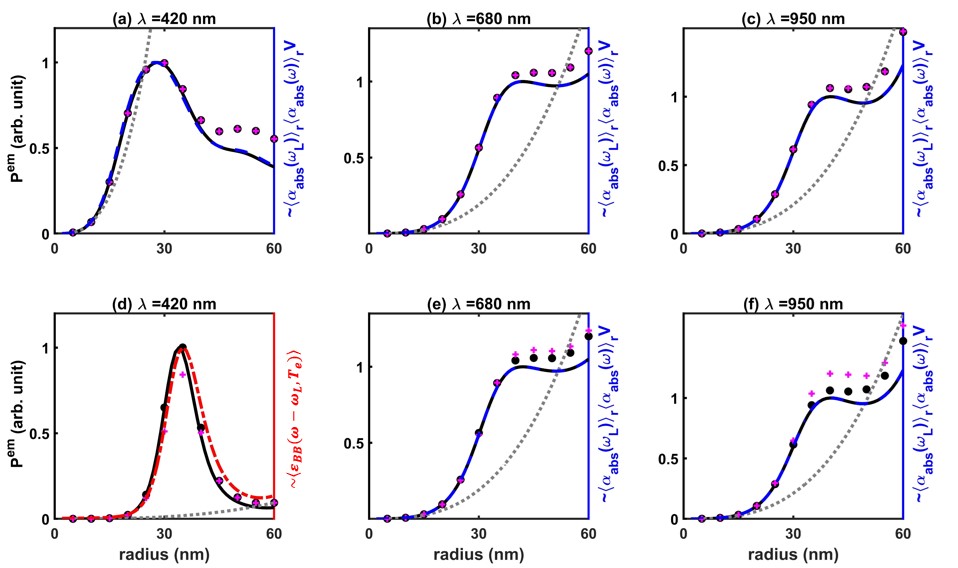}
\caption{ (Color online) The total PL (Eq.~(\ref{Eq:P_em_main}); circles) and its homogenized approximation (Eq.~(\ref{Eq:PL_CW_sph}); black continuous line) calculated for spheres under CW illumination with $\lambda_L = 488$ nm and emission wavelengths of $\lambda = 420$ nm ((a) and~(d)), $\lambda = 680$ nm ((b) and~(e)) and $\lambda = 950$ nm~((c) and~(f)). For these calculations, $p_{sat} \sim 1.2 \times 10^{25}$  W/m$^3$ and $I_{in}$ used for the calculation are~(a)-(c) $2.5$ kW$/$cm$^2$ and~(d)-(f) $0.25$ MW$/$cm$^2$. The magenta crosses show the PL calculation when the thermo-optic effect is included. The dashed blue lines represent $a^3\langle\alpha_{abs}(\lambda_L)\rangle_r\langle\alpha_{abs}(\lambda)\rangle_r$ and the red dash-dot lines represent $\langle \e_{BB}\rangle$. The gray dotted lines are an $a^3$ fit to the calculation for $a < 10$ nm. Frequency and wavelength are used interchangeably, as convenient. } \label{Fig.PL_488nm_sphere}
\end{figure}

Fig.~\ref{Fig.PL_488nm_sphere} illustrates the size-dependence of the total PL (as per Eq.~(\ref{Eq:P_em_main})) for an excitation wavelength close to resonance ($\lambda_L = 488$ nm). The intensities used for the calculations are chosen such that at the lower intensity ($I_{in} = 2.5 kW/cm^2$), the temperature of each sphere is close to room temperature and at the higher intensity ($I_{in} = 0.25 MW/cm^2$) the maximum temperature does not exceed $500 K$ (for which sintering and damage may start to occur). The calculations are conducted at three different emission wavelengths. The calculated emission at $\lambda = 420$ nm (aSE; Fig.~\ref{Fig.PL_488nm_sphere}(a)) displays a peak at $a = 28$ nm when excited by the low intensity. When excited at a higher intensity (Fig.~\ref{Fig.PL_488nm_sphere}(d)), the peak slightly red-shifts to $a = 34$ nm and becomes more distinct. For the emission at $\lambda = 680$ nm (SE; Fig.~\ref{Fig.PL_488nm_sphere}(b) and~(e)), the PL increases monotonously until $a \approx 42$ nm, decreases until $a = 51$ nm and then rises again. The PL at $\lambda = 950$ nm, (lower frequency SE; Fig.~\ref{Fig.PL_488nm_sphere}(c) and~(f)) exhibits a similar behavior, except for a higher slope beyond $a = 50$ nm.

As already noted above, the expression for PL~(\ref{Eq:P_em_main}) can be further simplified for a sphere. As discussed in Section~\ref{sec.T_sphere}, the temperature is fairly uniform in the sphere and the space-dependence of $I_e$ in Eq.~(\ref{Eq:P_em_main}) originates only from that of $\alpha_{abs}$ and $\delta_E$, as indicated by Eq.~(\ref{Eq:I_e}). Further, we rely on the calculations of the mean free path of the non-thermal electrons~\cite{hot_es_Atwater} to justify the neglect of spatial broadening of the non-thermal electron distribution. These results motivate the use of the analytic result~(\ref{eq:Ie_guess}) also for non-uniform fields, i.e., to approximate the level of non-thermal electron population $\delta_E$ by simply summing over its local value across the particle volume. Accordingly, $\delta_E$ (\ref{eq:delta_E}) can be expressed in terms of the spatially averaged, $\langle\alpha_{abs}({\bf r};\omega_L)\rangle_r$. With this consideration, $I_e$~(\ref{Eq:I_e}) becomes independent of position. The volume integration over $\alpha_{abs}({\bf r}; \omega)$ is now replaceable by the absorption cross-section of the sphere, which in turn, can be expressed as $\sigma_{abs}(\omega) \sim a^3 \langle \alpha_{abs}({\bf r};\omega)\rangle_r$. Thus, Eq.~(\ref{Eq:P_em_main}) simplifies to
\begin{eqnarray}\label{Eq:PL_CW_sph}
dP^{em}(\omega,\omega_L)
&\sim& \omega^2 a^3 \langle \alpha_{abs}({\bf r};\omega)\rangle_r \bigg[2
{\frac{I_{in}}{p_{sat}(\omega_L)}}\langle\alpha_{abs}({\bf r};\omega_L)\rangle_r\langle\e_{BB}(\omega - \omega_L;T_e)\rangle_r \nn \\
&&+ \bigg({\frac{I_{in}}{p_{sat}(\omega_L)}}\bigg)^2\langle\alpha_{abs}({\bf r};\omega_L)\rangle_r^2 \langle\e_{BB}(\omega - 2 \omega_L;T_e)\rangle_r\bigg] d\omega d\Omega.
\end{eqnarray}

Using this expression, we can determine the total PL solely via the absorption cross-section density evaluated at the pump frequency ($\langle \alpha_{abs}({\bf r};\omega_L) \rangle_r$, Fig.~\ref{Fig.alpha_and_T_sphere}(b)) and emission frequency ($\langle \alpha_{abs}({\bf r};\omega) \rangle_r$, Fig.~\ref{Fig.alpha_vs_WL}(a)), as well as the electron temperature (Fig.~\ref{Fig.alpha_and_T_sphere}(c)). As seen in Fig.~\ref{Fig.PL_488nm_sphere}(a)-(c), the qualitative behaviour of the PL (Eq.~(\ref{Eq:P_em_main})) and its approximation (Eq.~(\ref{Eq:PL_CW_sph})) is nearly identical except for the former being slightly higher for $a > 30$ nm~\footnote{\label{foot_sphere}This can be shown analytically via the Mie solution; the proof of this behaviour for the case of layers is shown in footnote~\ref{foot_layer}. }.

While the computational simplification associated with Eq.~(\ref{Eq:PL_CW_sph}) is minor, its more significant contribution is to allow us to pinpoint the origins of the behaviour observed in Fig.~\ref{Fig.PL_488nm_sphere}. Specifically, the PL in Fig.~\ref{Fig.PL_488nm_sphere}(a)-(c) pertains roughly to room temperature for all sphere sizes, making the Bose function ($\e_{BB}(\omega - \omega_L)$) essentially size-independent. In this scenario, the PL varies with sphere size as the product of $\langle \alpha_{abs}({\bf r};\omega_L) \rangle_r$ (at the absorption wavelength), $\langle \alpha_{abs}({\bf r};\omega)\rangle_r$ (at the emission wavelength) and the volume, $V$. For spheres of radius $a < 10$ nm (i.e., safely within the quasi-static regime), $\langle \alpha_{abs}\rangle_r$ has a very weak dependence on the volume (see Fig.~\ref{Fig.ACS_vs_radius}) so that the PL scales as $a^3$. {\XYZ For spheres with sufficiently large radii with respect to the skin depth, the absorption is dominated by the surface layer of the particle, so that $\langle \alpha_{abs}\rangle_r \sim 1/a$. Consequently, the PL scales linearly with $a$ ($\sim a^3 \frac{1}{a} \frac{1}{a}$). For sphere sizes between these two extremes,} the sub-volume scaling of the absorption and temperature, along with variations in $\langle\alpha_{abs}\rangle_r$ with particle size owing to the sphere's modal response (see~\onlinecite{Un-Sivan-size-thermal-effect} and Fig.~\ref{Fig.alpha_and_T_sphere}(b)) result in the observed deviation from the volume scaling, namely, the peaks and troughs in the PL shown in all subplots of Fig.~\ref{Fig.PL_488nm_sphere}. Specifically, the non-monotonic behaviour in the PL observed for the larger sizes are due to the combined effect of the system tuning into and then out of the dipole resonance, and then tuning into the quadrupole mode. In that respect, considering the PL as being a volume effect for small spheres, and as a surface effect for larger spheres, is a crude description, which misses the dominant effect of the modal structure {\XYZ but captures the baseline}.

Fig.~\ref{Fig.PL_488nm_sphere}(d)-(f) shows the PL for a higher intensity which results in a size-dependent temperature, varying from room temperature to $\sim 500 K$ (see Fig.~\ref{Fig.alpha_and_T_sphere}(c)). The higher varying temperature affects differently the different parts of the PL spectrum. First, it hardly affects the SE (Fig.~\ref{Fig.PL_488nm_sphere}(e)-(f)) which thus follows the same trend observed for the low-intensities, at least up to temperatures for which thermo-optic effects are still negligible~\cite{Gurwich-Sivan-CW-nlty-metal_NP,IWU-Sivan-CW-nlty-metal_NP} (see Section~\ref{sec.Thermo-optic-effect} below). Indeed, for SE, the non-thermal contribution from the $1$PA term dominates $I_e$ (see Eq.~(\ref{eq:Ie_guess}), Fig.~\ref{Fig.Schematic_of_Ie} and Fig.~\ref{Fig.Ie_vs_radius_sphere}), which is nearly temperature-independent since it originates from the non-thermal electron shoulders given in Eq.~(\ref{eq:delta_E}) (see SI Section S3 of~\onlinecite{Sivan-Dubi-PL_I}). In contrast, the aSE is exponentially sensitive to the temperature via the Bose function $\langle \e_{BB}(\omega - \omega_L;T_e)\rangle$ (see Fig.~\ref{Fig.Schematic_of_Ie}). As a result, the size-dependence of the PL is much stronger (see Fig.~\ref{Fig.PL_488nm_sphere}(d)), exhibiting superlinear scaling with the volume.

Deeper into the aSE regime, the $2$PA term becomes stronger than the 1PA term, see Fig.~\ref{Fig.Schematic_of_Ie}. To demonstrate this, we calculate the PL for a longer (off-resonant) excitation wavelength ($900$ nm, i.e., smaller $\omega_L$), see Fig.~\ref{Fig.PL_900nm_sphere}. In this case, while the near aSE (Fig.~\ref{Fig.PL_900nm_sphere}(b) and~(f)) grows as $\langle \alpha_{abs}({\bf r};\omega_L) \rangle_r$, for frequencies sufficiently deep into the aSE, the PL scales as $\sim \langle \alpha_{abs}({\bf r};\omega_L) \rangle_r^2$ at low intensity excitation (Fig.~\ref{Fig.PL_900nm_sphere}(a)) and as $\sim \langle \e_{BB}(\omega - 2 \omega_L)\rangle$ at high intensity illumination (Fig.~\ref{Fig.PL_900nm_sphere}(d)).

Finally, in order to show the generality of our approach, and to demonstrate that the behaviour we identify is not qualitatively affected by the choice of the exciting laser frequency (e.g., its position with respect to resonance), we replicate the results of Fig.~\ref{Fig.PL_488nm_sphere} (for which the illumination is on the red-side of the plasmon resonance) in Fig.~\ref{Fig.PL_488nm_n-1.8} and Fig.~\ref{Fig.PL_488nm_n-2.2} for backgrounds with higher refractive indices. This results in a red-shift of the resonances and introduces the higher-order modal response to the PL emission, particularly, for the larger spheres. However, the qualitative behaviour is similar to that with the lower background refractive index.

\subsubsection{The thermo-optic effect}\label{sec.Thermo-optic-effect}

Illumination at high intensity can induce significant changes in the temperature of the spheres, consequently altering their permittivity via the so-called thermo-optic effect~\cite{Boyd-book,Sivan-Chu-high-T-nl-plasmonics,korean_Chu_TO_nlty}, and as a result, modify the absorptivity (hence, the emissivity) and finally the PL. To analyze this effect, we assume that the permittivity of the metal and the thermal conductivity of the surrounding medium depend linearly on the change of temperature, compute the thermo-derivative of the metal permittivity from ellipsometry data~\cite{Shalaev_ellipsometry_silver} and set the thermo-derivative of the thermal conductivity of the surrounding medium to $1.3 \times 10^{-4}$ Wm$^{-1}$K$^{-2}$, as in~\onlinecite{IWU-Sivan-CW-nlty-metal_NP}; the temperature of the nanoparticle is then calculated using the method described by Un \textit{et al.}~\cite{IWU-Sivan-CW-nlty-metal_NP}.

Overall, as the intensity (hence, temperature) increases, the real part of the permittivity $(\epsilon_m')$ becomes more negative, while the imaginary part $(\epsilon_m'')$ increases to higher positive values at most frequencies. This results in a blue shift of the resonance peak and a broadening of its spectral width, leading to enhanced absorption at off-resonance frequencies, see Fig.~\ref{Fig.abs_spec_TOE} and~\onlinecite{PT_Shen_ellipsometry_gold,Shalaev_ellipsometry_gold,Shalaev_ellipsometry_silver,Sivan-Chu-high-T-nl-plasmonics,korean_Chu_TO_nlty}. As a result, $\langle \alpha_{abs}({\bf r};\omega) \rangle_r$ slightly decreases at $488$ nm and slightly increases at $900$ nm, as seen in {\XYZ Fig.~\ref{Fig.alpha_and_T_sphere}(b)}. The decrease in the former case leads to a reduction in the temperature for the hottest spheres (around $a = 30$ nm), and hence, to the reduction of the aSE observed in Fig.~\ref{Fig.PL_488nm_sphere}(d). However, the SE PL is hardly affected by the temperature. Thus, the increase in $\langle \alpha_{abs}({\bf r};\omega) \rangle_r$ at longer wavelengths results in a slight increase of the PL, as observed in Fig.~\ref{Fig.PL_488nm_sphere}(e) and~(f). For excitation at $900$ nm, the temperature of the spheres (especially for $a > 30$ nm) increases slightly compared to the value calculated for the linear case, leading to a slight increase in the PL for aSE, as shown in Fig.~\ref{Fig.PL_900nm_sphere}(d)-(e). The thermo-optic increase of $\langle \alpha_{abs}({\bf r};\omega_L) \rangle_r$ and $\langle \alpha_{abs}({\bf r};\omega) \rangle_r$ leads to a slight increase in SE, as observed in Fig.~\ref{Fig.PL_900nm_sphere}(f).


\subsubsection{Comparison to single particle experiments}\label{subsub:exp_sphere}

\begin{figure}[htp]
\includegraphics[width=17cm]{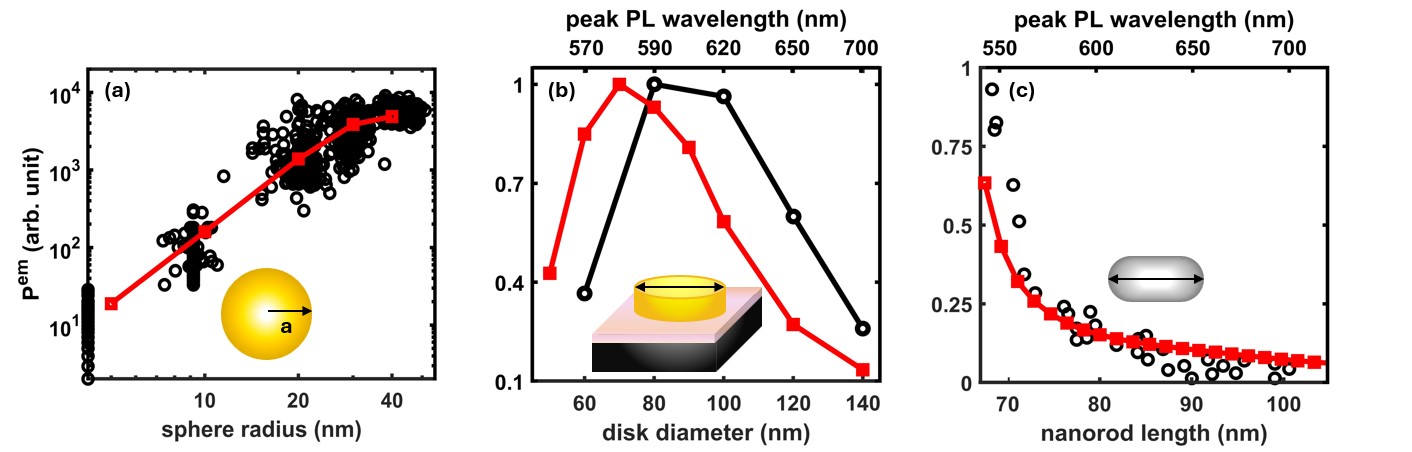}
\caption{ (Color online) (a) Normalized PL from gold spheres on a glass substrate immersed in glycerol with $\lambda_L = 514$ nm with emission integrated over the range of $560-640$ nm, as measured by Gaiduk \textit{et al.}~\cite{Orrit_QY_metal_PL_2011} \!\!\!. (b) PL spectrally-resolved peak positions and intensities vs. diameter of gold disks of height $30$ nm at $\lambda_L = 532$ nm as measured by Hu \textit{et al.}~\cite{Joel_Yang_ACS_Nano_2012}\!\!\!. (c) Spectrally-integrated PL from silver nanorods of diameter $23$ nm on a quartz substrate, at $\lambda_L = 532$ nm as measured by Lin \textit{et al.}~\cite{intraband_PL_Xiamen}\!. In all panels, experimental results and theoretical calculations using Eq.~(\ref{Eq:P_em_main}) are depicted by black circles and red squares, respectively, with corresponding solid lines added for visual guidance. The corresponding calculated PL peak positions are indicated on the top axis of (b) and (c). }
\label{Fig.Exp_sphere}
\end{figure}

Our theoretical predictions can be compared to various single particle PL measurements. First, Fig.~\ref{Fig.Exp_sphere}(a) shows that our calculations align with the experimental observations of the normalized SE PL (spectrally integrated over the range of $560$ nm to $640$ nm) measured from single gold nanospheres of different sizes, as reported by Gaiduk \textit{et al.} (see Fig.~S5 of~\onlinecite{Orrit_QY_metal_PL_2011}). {\XYZ The experimental results are presented as they appear in the referenced paper (in arbitrary units), and the corresponding theoretical calculations are scaled by a constant normalization factor to enable quantitative comparison with the experimental data. In this experiment,} the illumination intensity was adjusted to keep the temperature rise relatively low (approximately below $20$ K). The PL is calculated using Eq.~(\ref{Eq:P_em_main}) with the refractive index of gold obtained from~\onlinecite{optic_constant_noble_metals}, and those of glass and glycerol being $1.45$ and $1.47$, respectively. The normalized PL roughly scales with the volume until $a \sim 30$ nm, but seems to saturate upon further increase in size; assuming this is the onset of a decrease, this is in line with our prediction.

Size-dependence of the PL from larger particles was studied by Hu \textit{et al.}~\cite{Joel_Yang_ACS_Nano_2012} who demonstrated that the spectral peak of the PL from gold nano-disks on a SiO$_2$-coated absorptive (silicon) substrate initially increases, reaches a maximum, and then decreases significantly with further increases in size. The absorption cross-section densities and the $T_e$ of the disks are calculated using COMSOL, considering normal incidence and emission perpendicular to the substrate, with the permittivity data for gold and Si taken from~\onlinecite{optic_constant_noble_metals} and~\onlinecite{Refractive_index_of_Si_Aspnes}, respectively. Calculations based on Eq.~(\ref{Eq:P_em_main}) exhibit a qualitatively similar trend (even if at a slight spectral shift).

Lastly, our theoretical prediction of PL are compared with the PL measurements by Lin {\em et al.}~\cite{intraband_PL_Xiamen}. This work showed the spectrally-integrated PL from single nanorods of different lengths with constant diameter along their main axis. In contrast to the nanospheres (and in similarity to films, shown later), the PL monotonically decreases with nanorod length. These results are reproduced in Fig.~\ref{Fig.Exp_sphere}(c) along with the theoretical prediction, i.e.,~the spectral integration of Eq.~(\ref{Eq:P_em_main}). The optical response is calculated in COMSOL, where the substrate refractive index is $n = 1.45$ and the permittivity of the silver rods is taken from~\onlinecite{optic_constant_noble_metals}. The temperature is estimated with an approximate expression~\cite{thermo-plasmonics-basics} and remains close to the ambient temperature (0.1\% variation) for the laser illumination intensity considered ($I_{in} = 3$ mW/cm$^2$). Good qualitative agreement is obtained for nanorods longer than $72$~nm, however, our prediction underestimates the PL for shorter nanorods, thus providing only a qualitative match for this regime (specifically, in a 30 nm wide wavelength range close to the pump). A similar observation was recently reported above the pump wavelength {\XYZ for Au rods~\cite{Pelous-Greffet}; in that case, the discrepancy between theory and measurements was ascribed to interband transitions.}

Remarkably, while our theory~(\ref{Eq:P_em_main}) includes only emission events occurring in the conduction band, it successfully predicts the dependence of the emission on the particle size even in spectral regimes where interband emission events should occur (specifically, for the spheres and nano-disks (Fig.~\ref{Fig.Exp_sphere}(a)-(b), respectively). This implies that, in these cases, interband transitions modify the emission features in a modest quantitative manner, but not qualitatively. We expect that inclusion of such transitions may yield a better quantitative match to the experimental results{\XYZ, due to a better match of the permittivity and non-thermal electron distribution}.

\subsection{Films under continuous illumination} \label{sub:layers}
We now turn to study the PL from a thin film of a Drude metal using the approach used above. Specifically, we examine a thin silver film on a glass substrate illuminated at normal incidence (hence, no dependence on polarization $pol'$, and $k_z = |\bf k'|$, where $k_z$ is the component of the wavenumber {\bf k} normal to the surface of the film) by a focused Gaussian CW beam with an intensity profile $I_{in}(\rho,\omega_L) = I_0 e^{- 2 \rho^2/b^2}$ as depicted in Fig.~\ref{Fig.schematic_T_film}(a).

\subsubsection{Determination of the electron temperature}

\begin{figure}[htp]
\centering
\includegraphics[width=15cm]{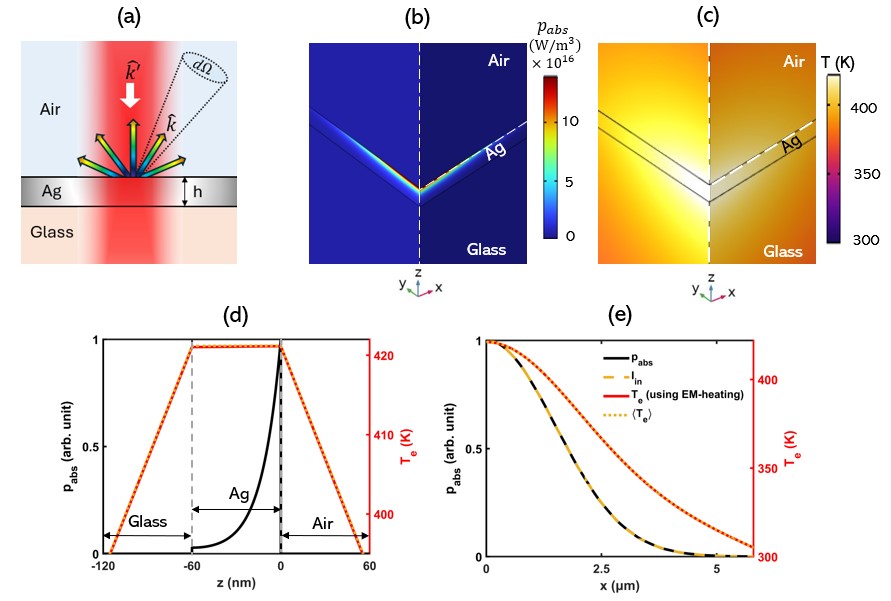}
\caption{ (Color online) (a) Schematic illustration of the emission from an Ag film on a glass substrate illuminated from above by a focused Gaussian beam. (b) The absorbed power density calculated for a $60$ nm film due to continuous wave illumination using the electromagnetic heating module of COMSOL Multiphysics (shown on a quarter domain). (c) The corresponding calculated temperature distribution. (d) Vertical (along the $z$-axis) and~(e) transverse ($x$-axis, $z = 0$) cross-sections of the absorbed power density and temperature. The red-solid and yellow dotted lines in~(d) and~(e) represent $T_e$ calculated using  Eq.~(\ref{Eq_T_Steady}), considering the actual distribution of $\alpha_{abs}({\bf r};\omega_L)$ and the position averaged $\langle\alpha_{abs}({\bf r};\omega_L)\rangle_z$ respectively. The yellow dashed line in~(e) represents the distribution of the incident intensity along the $x$-axis. The peak intensity and radius of illumination are $2$ MW/cm$^2$ and $3~\mu$m, respectively. } \label{Fig.schematic_T_film}
\end{figure}

The absorbed power density and temperature distribution calculated by solving the heat equation~(\ref{Eq_T_Steady}) using the electromagnetic heat module of COMSOL Multiphysics are presented in Figs.~\ref{Fig.schematic_T_film}(b)-(c), respectively. The absorbed power density decreases exponentially along the illumination direction ($z$-axis) due to the short penetration depth (Fig.~\ref{Fig.schematic_T_film}(d)). However, along the radius it exhibits nearly the same distribution as that of the incident intensity, as shown in Fig.~\ref{Fig.schematic_T_film}(e).

In contrast, the temperature variation across the film's thickness is minimal (see Fig.~\ref{Fig.schematic_T_film}(d)) due to the high thermal conductivity of the metal; for the same reason, along the radius, the temperature shows a Gaussian profile with a full width at half maximum (FWHM) greater than that of the incident beam, as illustrated in Fig.~\ref{Fig.schematic_T_film}(e). 

Generically, $\alpha_{abs}({\bf r};\omega_L, pol',\hat{k'})$ is uniform in the plane of the film (i.e., it is independent of $\rho$ and $\phi$) and varies only along the thickness of the film (i.e., it is only $z$-dependent). The average value of $\alpha_{abs}(z,\omega_L, pol',\hat{k'})$ along the thickness can be derived from the experimentally measurable quantity, the absorptance $A(\omega, pol',\hat{k'})$, as
\begin{equation} \label{eq:absorptance}
\langle\alpha_{abs}({z},\omega_L, pol',\hat{k'})\rangle_z = \frac{1}{h}\int_h \alpha_{abs}({z},\omega_L, pol',\hat{k'})dz = \frac{A(\omega_L, pol',\hat{k'})}{h}.
\end{equation}
$\alpha_{abs}({\bf r};\omega_L, pol',\hat{k'})$ and its average can also be calculated using the transfer matrix method~\cite{transfermatrix1990Ishida}. For normal incidence, this quantity is independent of polarization.

Since the variation of temperature along the film thickness is negligible (see Figs.~\ref{Fig.schematic_T_film}(c)-(d)), the temperature of the film can be calculated by using the $z$-averaged cross-section density $\langle \alpha_{abs}(z,\omega_L, pol',\hat{k'})\rangle_z$ in Eq.~(\ref{Eq_T_Steady}). The temperature calculated for a $60$ nm Ag film with this approximation closely matches the temperature calculated based on the exact absorbed power density distribution, as shown in Fig.~\ref{Fig.schematic_T_film}(e). This also enables simplifying the expression for the PL in Eq.~(\ref{Eq:P_em_main}) by separating the integrations with respect to  $z$ and $\rho$, as discussed in more detail in Section~\ref{sec_PL_films}.

\begin{figure}[htp]
\centering
\includegraphics[width=13cm]{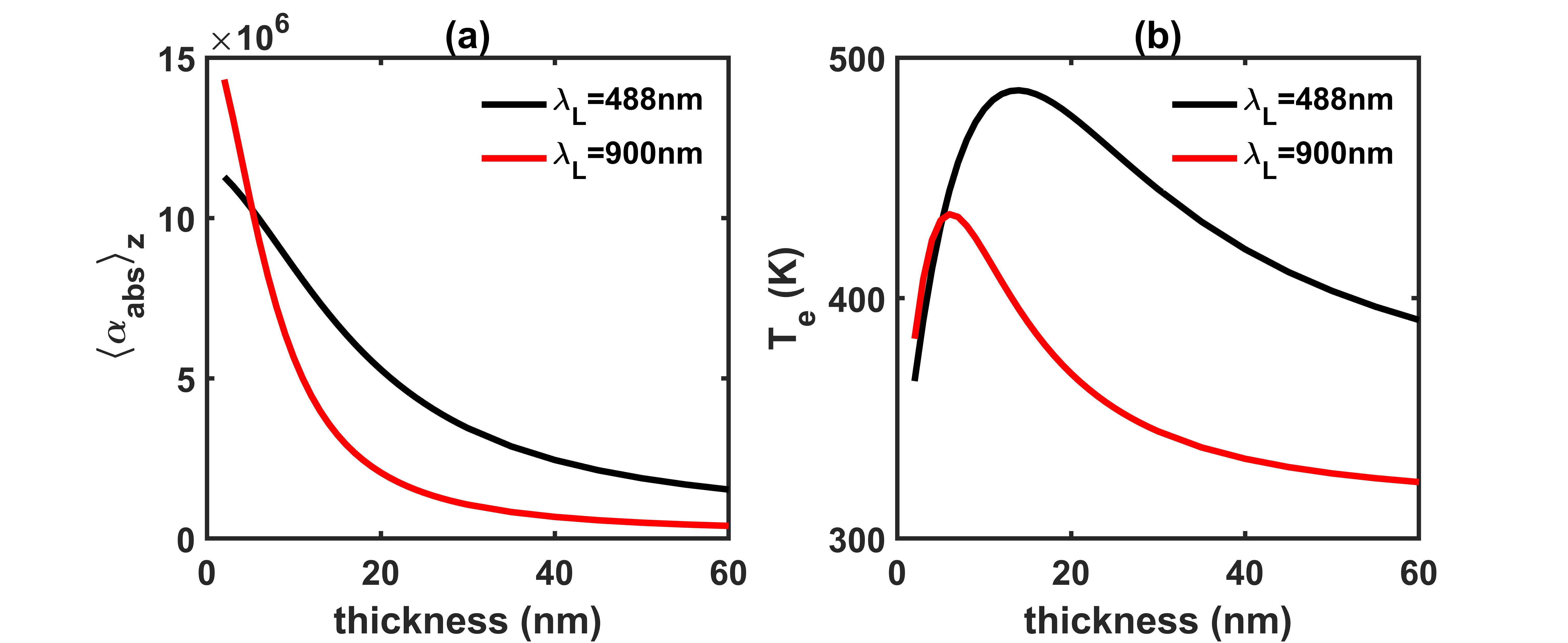}
\caption{ (Color online) (a) $\langle \alpha_{abs}\rangle_z$ and~(b) the (electron) temperature of Ag films at the center of the illumination by a CW Gaussian beam with central intensity {\XYZ $I_0 = 1.5$ MW/cm$^2$} and $b = 3\mu$m, with $\lambda_L = 488$ nm (black) and $\lambda_L = 900$ nm (red).}  \label{Fig.alpha_T_film}
\end{figure}
The size-dependence of $\langle \alpha_{abs}(z;\omega_L)\rangle_z$ for normal incidence at $\lambda_L = 488$ nm and $900$ nm are shown in Fig.~\ref{Fig.alpha_T_film}(a). $\langle \alpha_{abs}(z,\omega_L)\rangle_z$ decreases with increasing film thickness for both wavelengths, with this decay occurring more rapidly at the longer wavelength. This trend can be explained by the distribution of the electric field within the metal film. Typically, the electric field is strongest near the surface of the metal film and exhibits an exponential decay as it penetrates deeper into the material. Thus, as the film thickness increases, the average field, hence, absorptance density $A / h$ decreases. This decay is more rapid for $\lambda = 900$ nm due to its shorter penetration depth ($= 21$ nm) compared to the penetration depth of $26$ nm at $\lambda_L = 488$ nm as shown in Fig.~\ref{Fig.alpha_T_film} and Fig.~\ref{Fig.alpha_vs_WL}(b).

Unlike the case of spheres, the size-dependence of the film temperature differs from $\alpha_{abs}$. The temperature at the center of the illumination spot ($\rho = 0$) as a function of film thickness is depicted in Fig.~\ref{Fig.alpha_T_film}(b). Maximal temperature is reached at $h = 14$ nm and $6$ nm, corresponding to the short and long wavelengths, respectively. This peak temperature can be explained from Fourier's law of heat transfer (solution of the divergence in Eq.~(\ref{Eq_T_Steady})), according to which, $T_e \propto h \langle\alpha_{abs}(z;\omega_L)\rangle_z$~\footnote{Solving the divergence in Eq.~(\ref{Eq_T_Steady}) simply along z-axis, we obtain $-\kappa \nabla T_e(z,\rho = 0) \sim h\langle\alpha_{abs}(z;\omega_L)\rangle_z I_{in}(\rho = 0;\omega_L)$. Since the temperature inside the film remains nearly constant along z-axis, and decreases linearly to $T_h$ at a distance $\Delta z$ outside the film, $T_e(z,\rho = 0) - T_h \sim \frac{\Delta z}{\kappa} h \langle\alpha_{abs}(z;\omega_L)\rangle_z I_{in}(\rho = 0;\omega_L)$.}. Thus, with increasing thickness, the temperature initially increases. However, due to the sharp decline of $\langle \alpha_{abs}({\bf r};\omega_L)\rangle_r$, with thickness, the temperature eventually decreases, resulting in a peak.

\subsubsection{Determination of the PL}\label{sec_PL_films}
\begin{figure}[htp]
\centering
\includegraphics[width=16.5cm]{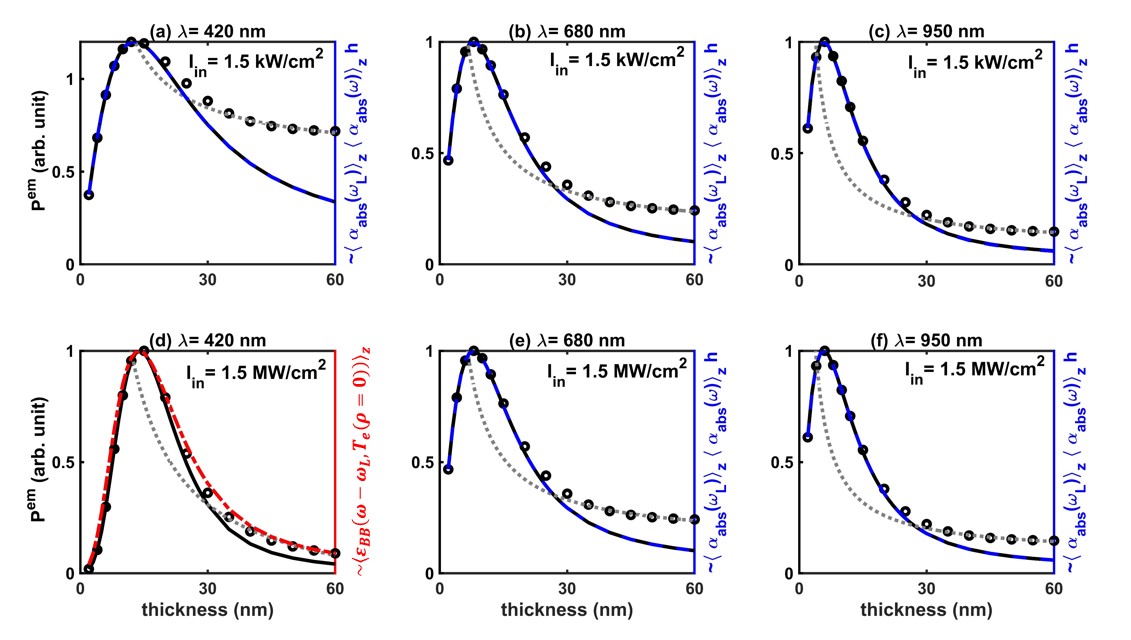}
\caption{ (Color online) The total PL calculated for the films using Eq.~(\ref{Eq:P_em_main}) (circles) and that using uniform field approximation (black continuous line) under CW illumination with $\lambda_L = 488$ nm and emission wavelengths of $\lambda = 420$ nm ((a) and~(d)), $\lambda = 680$ nm ((b) and~(e)) and $\lambda = 950$ nm ((c) and~(f)). For these calculations, $p_{sat} \sim 1.2 \times 10^{25}$  W/m$^3$ and $I_0$ values are (a)-(c) {\XYZ $1.5$ kW$/$cm$^2$} and~(d)-(f) {\XYZ $1.5$ MW$/$cm$^2$}. The dashed blue lines represent $\langle \alpha_{abs}(\omega_L)\rangle_z \langle \alpha_{abs}(\omega)\rangle_z h$ and the red dash-dot lines represent $\langle \e_{BB}\rangle$ at $\rho = 0$. {\XYZ The gray dotted lines represent a $1/h$ fit to the calculations for $h >30$ nm.} } \label{Fig_PL_Film_Ag_488nm}
\end{figure}

Fig.~\ref{Fig_PL_Film_Ag_488nm} presents the predicted PL obtained using Eq.~(\ref{Eq:P_em_main}) for Ag films of thickness ranging from $2$ nm to $60$ nm. For this calculation, we consider normal illumination at a wavelength of $488$ nm, with the emitted PL collected within a solid angle $d\Omega$ positioned directly above the illumination point. As above, calculations are performed at two incident intensities: one for which the films remain roughly at room temperature {\XYZ ($I_0 = 1.5 kW/cm^2$)} and the other to elevate their temperature to {\XYZ $486$ K} {\XYZ ($I_0 = 1.5$ MW/cm$^2$)} as shown in Fig.~\ref{Fig.alpha_T_film}(b). At room temperature, the PL at the three emission wavelengths ($420$ nm, $680$ nm and $950$ nm) shows peaks at film thicknesses of $13$ nm, $8$ nm and $6$ nm, respectively. Similar to the case of the spheres, when the peak illumination intensity is increased to {\XYZ $I_0 = 1.5$ MW/cm$^2$}, the trends of SE remain unchanged so that Figs.~\ref{Fig_PL_Film_Ag_488nm}(e)-(f) are hardly distinguishable from Figs.~\ref{Fig_PL_Film_Ag_488nm}(b)-(c). However, the peak corresponding to aSE becomes narrower.

Similar to the analysis of the PL from particles above, the PL can be again approximated by replacing $\alpha_{abs}(z)$ by its average value along the thickness, $\langle \alpha_{abs}(z) \rangle_z$ as
\begin{equation}
\begin{split}\label{Eq_PL_film_avg}
dP^{em}(\omega,\omega_L,\hat{k}) &\sim
\sum_{pol = s,p} \frac{\omega^2}{4 \pi^2 c^2} \bigg[\frac{1}{p_{sat}(\omega_L)}\bigg(h~\langle\alpha_{abs}({z},\omega_L)\rangle_z
\langle\alpha_{abs}(z,\omega)\rangle_z \\
&\int_{\rho = 0}^\infty 2 I_{in}(\rho,\omega_L)\langle\e_{BB}(\omega - \omega_L;T_e(\rho))\rangle_z \rho d\rho \bigg) \\
&+ \frac{1}{p_{sat}^2(\omega_L)}\bigg(h \langle\alpha_{abs}^2({z},\omega_L)\rangle_z
\langle\alpha_{abs}({z},\omega)\rangle_z\\
&\int_{\rho = 0}^\infty I_{in}^2(\rho,\omega_L)\langle\e_{BB}(\omega - 2 \omega_L;T_e(\rho))\rangle_z \rho d\rho \bigg) \bigg] d\omega d\Omega.
\end{split}\end{equation}
Again, the calculations~(\ref{Eq_PL_film_avg}) closely match the results with the exact calculations (Eq.~(\ref{Eq:P_em_main})), with the SE (Fig.~\ref{Fig_PL_Film_Ag_488nm}(b)-(c),(e)-(f)) and low temperature aSE (Fig.~\ref{Fig_PL_Film_Ag_488nm}(a)) being well approximated by $\langle \alpha_{abs}(z,\omega_L)\rangle_z \langle \alpha_{abs}(z,\omega)\rangle_z h$ for films up to $h \sim 20$ nm, after which the approximation falls below the exact results\footnote{\label{foot_layer}For films thicker than $30$ nm, the PL calculated with Eq.~(\ref{Eq_PL_film_avg}) is less than that calculated with Eq.~(\ref{Eq:P_em_main}) because $\frac{\int_0^h {\alpha_{abs}(z,\omega_L)\alpha_{abs}(z,\omega)dz}}{h\langle\alpha_{abs}(z,\omega_L)\rangle_z\langle\alpha_{abs}(z,\omega)\rangle_z}\sim{\frac{2n''(\omega)k(\omega) n''(\omega_L)k(\omega_L)}{n''(\omega)k(\omega) + n''(\omega_L)k(\omega_L)}} > 1$ for $n'' > 1$.}.

{\XYZ As for particles, we can now deduce the general scaling of the PL with the size from Eq.~(\ref{Eq_PL_film_avg}). Specifically, for very thin films ($h < 5$ nm), $\langle \alpha_{abs}(z)\rangle_z$ decays relatively slowly. As a result, due to the presence of the $h$  term in the PL expression in Eq.~(\ref{Eq_PL_film_avg}), the PL in this regime exhibits sublinear growth. For thicker films ($h > 30$ nm), $\alpha_{abs}(z) \rangle_z \sim 1/h$ due to the fact that total absorption becomes almost independent of film thickness ($\int \alpha_{abs} dz = Const$). Consequently, the PL scales as $\frac{1}{h} \frac{1}{h} h = 1 / h$, as demonstrated in Fig.~\ref{Fig_PL_Film_Ag_488nm}(c) and (f) and. In the intermediate region, a peak appears.} The stronger decay of $\langle \alpha_{abs}(z) \rangle_z$ at longer wavelengths causes the PL peak to shift slightly towards smaller values of $h$ {\XYZ (see Fig.~\ref{Fig_PL_Film_Ag_488nm}(d),~(f) and Fig.~\ref{Fig_Film_Ag_900nm})}. As seen in Fig.~\ref{Fig_PL_Film_Ag_488nm}(d), for higher temperatures, the aSE undergoes exponential amplification via the $\langle\e_{BB}(\omega - \omega_L;T_e(\rho))\rangle_z$ term, resulting in sharper peaks compared to those at the room temperature.

The PL calculated at the longer excitation wavelength of $900$ nm allows us to explore the PL behaviour deeper into the aSE regime, where the 2PA contribution takes over the 1PA contribution, see Figs.~\ref{Fig_Film_Ag_900nm}(a) and~(d). For this excitation wavelength, while the PL in the near aSE region (Fig.~\ref{Fig_Film_Ag_900nm}(b) and~(e)) can be explained using the same reasoning applied to $\lambda_L = 488$ nm illustrated in Fig.~\ref{Fig_PL_Film_Ag_488nm}(a) and~(d), in the deeper aSE regime, the PL varies as $\sim \langle \alpha_{abs}(z,\omega_L) \rangle_z^2$ (rather than $\langle \alpha_{abs}(z,\omega_L) \rangle_z$) at room temperature and as $\sim \langle \e_{BB}(\omega - 2 \omega_L, T_e(\rho))\rangle_z$ (rather than $\langle \e_{BB}(\omega -  \omega_L, T_e(\rho))\rangle_z$) at the high temperature excitation. It should be noted that, in the high-temperature 2PA aSE case shown in Fig.~\ref{Fig_Film_Ag_900nm}(d), $\langle \e_{BB}(\omega - \omega_L, T_e(\rho))\rangle_z$ overestimates the exact calculation~(\ref{Eq:P_em_main}) due to the additional dependence of the exact PL on $\langle \alpha_{abs}(z,\omega_L) \rangle_z^2$. %
As the temperature increases further, $\langle \e_{BB}(\omega -  \omega_L, T_e(\rho))\rangle_z$ provides a closer estimate of the exact PL.

{\XYZ Note that unlike the case of particles, the PL from the film decreases monotonically beyond the increase for small sizes. The reason for that is the absence of the resonances that occur for particles, into which the system is tuning in and out with the varying size. }

\subsubsection{Comparison to experiments} \label{subsub:exp_layers}
\begin{figure}[htp]
\centering
\includegraphics[width=12cm]{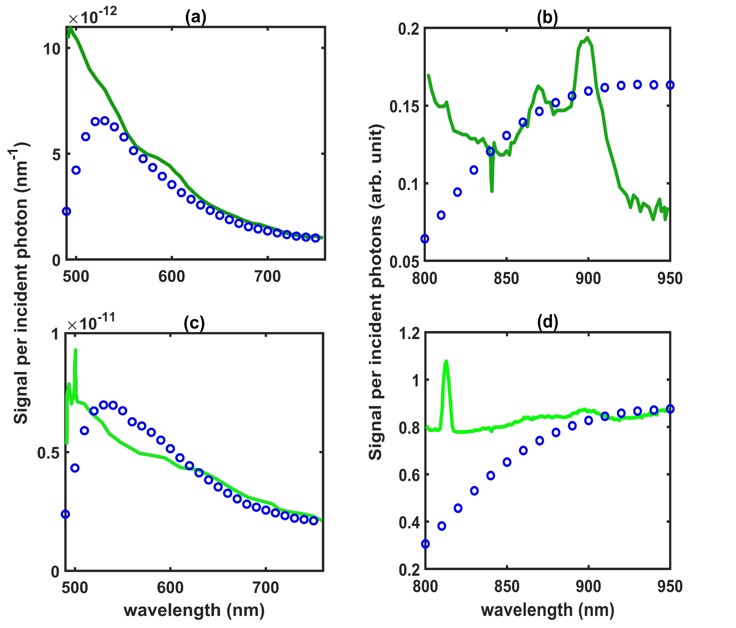}
\caption{{\XYZ (Color online) The recorded PL signal (number of emitted photons, solid lines) per incident photon per emission wavelength for film thicknesses (a)-(b) $h = 33.3$ nm and (c)-(d) $h = 13.4$ nm, taken from Fig.~3(a) and Fig.~S20 (a) of Bowman {\em et al.}~\cite{Abajo_Tagliabue_PL}\!\!, shown alongside our calculations (circles). Data of (a) and~(c) is given in units {\XYZ nm$^{-1}$} {\em multiplied by a constant factor of 8} for excitation wavelength of $\lambda_L = 488$ nm {\XYZ with $I_{in} = 0.162$ mW/$\mu$m$^2$} and the data of (b) and~(d) is given in {\XYZ arbitrary units multiplied by a constant factor of $6 \times 10^{13}$ for $\lambda_L = 785$ nm with $I_{in} = 1.87 $mW/$\mu$m$^2$}. In these calculations, the PL is integrated over a solid angle with the emission angle ranging from $0$ to $44^{\circ}$ corresponding to the numerical aperture of the objective lens used in the experiment.} }
\label{Fig_exp_Bowman}
\end{figure}

\begin{figure}[htp]
\centering
\includegraphics[width=12cm]{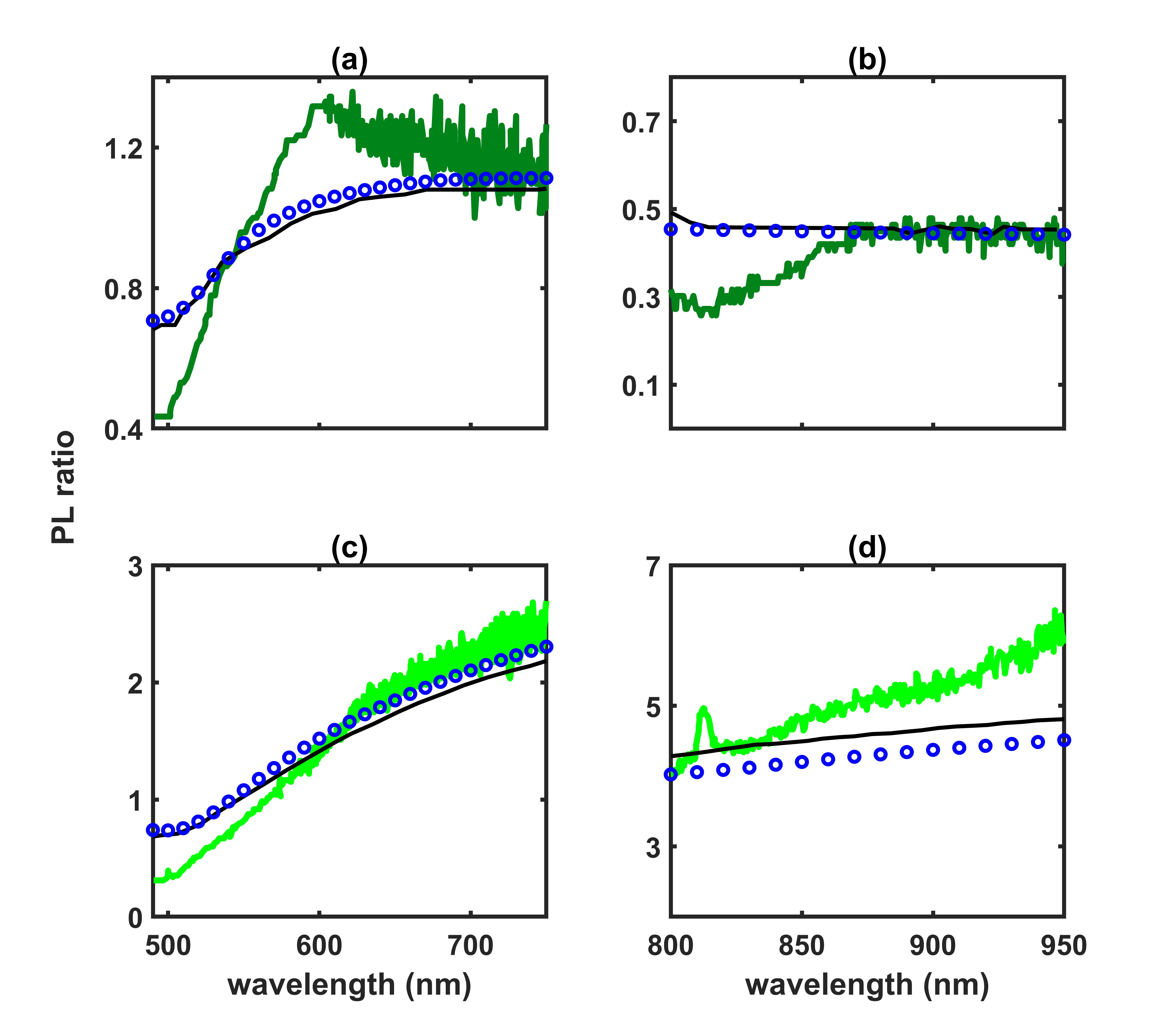}
\caption{(Color online) {\XYZ Corresponding PL ratios} experimental (colored solid lines) and simulation (black solid lines) results for the PL from golds films of thickness (a)-(b) $h = 33.3$ nm and (c)-(d) $h = 13.4$ nm {\XYZ to that of a thicker film ($h = 113$ nm for (a) and (c) and $h = 88$ nm for (b) and (d))}, taken from Fig.~3(c) and Fig.~S20(b) of Bowman {\em et al.}~\cite{Abajo_Tagliabue_PL}\!\! . They are compared to our calculation (Eq.~(\ref{Eq:P_em_main})) shown as blue circles. Panels (a) and~(c) show the ratio $\frac{\text{PL}(h)}{\text{PL}(h = 113\text{nm})}$ as a function of emission wavelength, using excitation conditions of $\lambda_L = 488$ nm, and panels (b) and~(d) display the normalized difference $\frac{\text{PL}(h) - \text{PL}(h = 88\text{nm})}{\text{PL}(h = 25.8\text{nm}) - \text{PL}(h = 88\text{nm})}$ at $\lambda_L = 785$ nm.}
\label{Fig_exp_Bowman_norm}
\end{figure}

As done above for single particles, we now compare the predictions of our simple analytic PL formula to recent experimental measurements. Specifically, we look at {\XYZ Figs.~3(a)-(c) and Figs.~S20(a)-(b)} of Bowman {\em et al.}~\cite{Abajo_Tagliabue_PL}, which presents PL measurements from single crystal Au films of various thicknesses. Only a few film thicknesses were studied, but data is available for several emission frequencies (rather than for single frequencies, as for particles). {\XYZ The PL recorded for the individual films is shown in Fig.~\ref{Fig_exp_Bowman}, alongside our calculations. To conform with the data of the original paper, our results for $\lambda_L = 488$ nm (Fig.~\ref{Fig_exp_Bowman}(a) and~(c)) are given in units of nm$^{-1}$ (as in Figs.~3(a)-(c) of~\onlinecite{Abajo_Tagliabue_PL}, multiplied by a constant factor of 8), while those for $\lambda_L = 785$ nm (Fig.~\ref{Fig_exp_Bowman}(b) and~(d)) are presented in arbitrary units (as in Fig. S20(a) of~\onlinecite{Abajo_Tagliabue_PL}, and are normalized using a method analogous to that employed for the experimental results of spheres shown in  Fig.~\ref{Fig.Exp_sphere}(a). For $488$ nm excitation, the calculated results tend to fall below the experimental data at shorter wavelengths. This discrepancy likely arises from the omission of interband contributions in our emission calculations.

In order to remove the impact of instrumental noise and a substrate-related Raman peak, we follow~\onlinecite{Abajo_Tagliabue_PL} and show
the PL {\XYZ ratio} spectra {\XYZ for the same films (i.e., the PL data / difference normalized by the data from a thick film) which are} presented in Fig.~3(c) and Fig.~S20(b) from Bowman {\em et al.}~\cite{Abajo_Tagliabue_PL} in Fig.~\ref{Fig_exp_Bowman_norm}(a) and (c) and Fig.~\ref{Fig_exp_Bowman_norm}(b) and~(d), respectively, along with our own theoretical prediction for films (Eq.~(\ref{Eq:P_em_main})). {\XYZ For the $488$ nm excitation, {\em despite} the differences observed at short wavelengths in Fig.~\ref{Fig_exp_Bowman}(a) and~(c), our calculations now closely match the theoretical results of Bowman {\em et al.}~\cite{Abajo_Tagliabue_PL} for all thicknesses studied (see results for the thicker films (see Fig.~\ref{Fig_exp_Bowman_thicker}). A similar match can be observed for $785$ nm excitation as well (see Fig.~\ref{Fig_exp_Bowman_norm}(b) and~(d))
.} In fact, our theoretical predictions match the more sophisticated theory of~\onlinecite{Abajo_Tagliabue_PL} even in the parts of the spectra where the latter does not match the experimental data well, leaving the discrepancies unexplained by either theoretical approach. This is remarkable, since the theory in~\onlinecite{Abajo_Tagliabue_PL} involved the summation of the individual dipole emissions, a treatment of the electron states as a discrete set in momentum space, and a detailed DFT-based calculation of the permittivity. Those we replaced by the local Kirchhoff Law, empiric permittivity data, and our simple analytic formula for the PL (Eq.~(\ref{Eq:P_em_main})) based on a continuous set of energy states, along with standard macroscopic electromagnetic and thermal calculations. Thus, overall, the quantitative match for the normalized data implies that the contribution of interband transitions to the PL exhibits a size dependence very similar to the intraband transitions -based prediction. 
}

\section{Discussion}\label{sec:Discussion}

In this work, we provide a unifying explanation for the dependence of the PL from metal nanostructures on their size. Specifically, using the experimentally-established expression for the non-equilibrium electron distribution in the conduction band~\cite{Dubi-Sivan,Dubi-Sivan-Faraday,Dubi-Sivan-MJs,Gabelli,Kumagai-ACS-phot-2023}, together with recent extension of the local Kirchhoff Law for light emission from metals characterized this non-equilibrium electron distribution~\cite{Pelous-Greffet}, we provide a simple analytic form for the (intensity, temperature and) size-dependence of the PL. In particular, we show that the size-dependence of the Stokes emission (SE) is determined primarily by the size-dependence of the absorption (= emission) cross-section density (aka emissivity density) but that the aSE is much more sensitive to the structure size at high illumination intensities due to the exponential dependence of the Bose function $\langle \e_{BB}(\omega,T_e) \rangle$ on the electron temperature $T_e$.

\begin{figure}[htp]
\centering
\includegraphics[width=12cm]{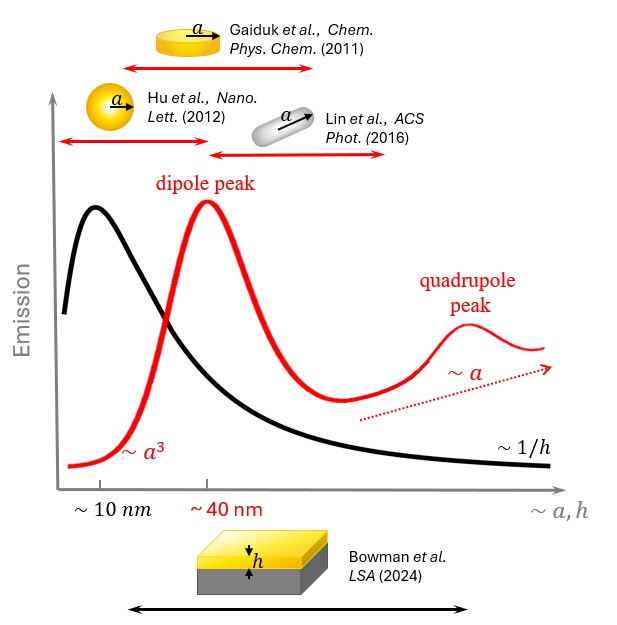}
\caption{\XYZ (Color online) Schematic summary of size dependence of the PL from the structures studied in this work.
\label{Fig_Schematic_summary} }
\end{figure}

{\XYZ This behaviour is illustrated schematically in Fig.~\ref{Fig_Schematic_summary}. As seen, generically, the PL from particles} and films grows for small sizes as the volume / thickness, respectively, and then decreases due to the decrease of the average local field (or equivalently, of the emissivity). The size for which maximal PL is attained is found to be larger for particles than for films ($\sim 30$ nm vs. $\sim 10$ nm, respectively, for illumination near resonance; the maximal size is higher for off-resonant illumination). This explains the seemingly contradicting trends observed {\XYZ in experiments} (e.g.,~\onlinecite{Feldman_QY_metal_PL,Orrit_QY_metal_PL_2011,Hecht_PRL-2019}) - 
they originate from the fact that measurements from thin films~\cite{Hecht_PRL-2019,Abajo_Tagliabue_PL} were so far performed only for films thicker than the size of maximal PL ($> 10$ nm), most likely, because thinner films tend to be discontinuous (i.e., below the percolation limit). A similar behaviour was observed for rods. 
{\XYZ Once advanced fabrication of sub 10nm thick films will be employed for (CW) PL studies (as, e.g., in~\onlinecite{Hecht_PRL-2019}), the quantum effects predicted in~\onlinecite{Abajo_Tagliabue_PL} could be identified. Notably, the trends described in Fig.~\ref{Fig_Schematic_summary} apply both for spectrally-resolved and spectrally-integrated PL; moreover, the trends will be qualitatively similar also for different wavelengths, for higher temperatures/intensities and for the aSE, in which cases, the sensitivity to the size will be greater.}

Our predictions were shown to match well a range of experimental results of SE, including for cases that formally lie beyond the limits of validity of the analysis, including non-spherical particles and their arrays. The quantitative discrepancies observed in some cases may originate in differences in the underlying permittivity data, in subtleties associated with the significant field non-uniformity in this case, or in missing elements needed for a complete description of the PL, e.g., the inclusion of emission originating from inelastic electron scattering\footnote{According to~\onlinecite{Baumberg-Aizpurua_Nat_comm_2024_inelastic}, this contribution dominates the emission near the pump frequency.}{\XYZ , interband transitions and the subtle QM effects highlighted in~\onlinecite{Abajo_Tagliabue_PL}.} Even better understanding of these discrepancies may be obtained by comparing to aSE measurements, which were so far done in the context of thermometry~\cite{Cahill_T_measure,Orrit-Caldarola_T_measure,Orrit-Caldarola_T_measure_transient,Kall_ACS_photonics_2018}, but not done systematically as a function of the structure size.

Our analytic approach provides a very simple alternative to the highly complicated rigorous (discrete $k$-space) calculations of the PL offered in some recent studies~\cite{Abajo_DTU_PL,Abajo_Tagliabue_PL,Baumberg-Aizpurua_Nat_comm_2024_inelastic}. In particular, the use of the extension of the local Kirchhoff Law to non-equilibrium electron distributions~\cite{Pelous-Greffet} allows circumventing many of the detailed electromagnetic emission calculations done in~\onlinecite{Abajo_Tagliabue_PL} and validates the approximations we made, e.g., the neglect of the energy-dependence of the matrix elements and electron density of states; this finding is in line with the good agreement obtained when matching our theoretical approach to measurement of currents in molecular and current junctions~\cite{Dubi-Sivan-MJs,Shalaev_Reddy_Science_2020,Gabelli,Kumagai-ACS-phot-2023}. Moreover, while the analytic expression we provide accounts for only intraband transitions, it is found to yield good qualitative and sometimes even quantitative agreement with experimental data from Au nanostructures in regimes for which interband emission was shown to be significant\footnote{In this context, the modelling of the contribution of interband transitions to the PL in~\onlinecite{Sheldon-nanoletters-2024} should be questioned, especially since it is based on an unjustified $100-$fold reduction of the electron-phonon coupling coefficient. }. 

Nevertheless, while both rigorous $k$-space approach and our approximate analytic approaches are capable of reproducing most aspects of the experimental observations, the current study highlights the ability of the more rigorous approaches to capture subtle aspects such as the determination of the conditions under which the emission is due to radiative recombination or inelastic scattering (see, e.g., discussions in~\onlinecite{Shen_RRS_vs_hot_L,Klein_RRS_vs_hot_L,Solin_Merkelo_RRS_vs_hot_L,Solin_Merkelo_RRS_vs_hot_L_reply_to_comment,Shen_RRS_vs_hot_L_comment,Baumberg_SERS_T_measure,Baumberg-Aizpurua_Nat_comm_2024_inelastic,Abajo_Tagliabue_PL}), {\XYZ emission from pre-scattered electrons~\cite{Abajo_Tagliabue_PL}, identifying subtle quantum mechanical effects that go beyond the standard quantum size effect (e.g.,~\onlinecite{Kreibig-book,Whetten_Quantum_size_effect_1997})\footnote{In comparison, the only quantum aspect of our theory is the use of the quantum version of the Boltzmann equation (see~\onlinecite{Dubi-Sivan}), while the electron states themselves (and thus, the permittivity) and the photon properties do not involve quantization. } such as the thickness dependence of the CW PL from atomic flat metal films (see Fig.~4a of~\onlinecite{Abajo_Tagliabue_PL}) and in the future, maybe even the high sensitivity to the number of atomic levels observed in~\onlinecite{Hecht_PRL-2019} for the PL due to pulsed illumination. }

While in the current work we focused on the PL from simple structures and noble metals, our approach can be applied also to more complicated structures for which the local electric field is even more non-uniform such as long rods~\cite{Bouhelier_PL_2019}, various particle dimers~\cite{metal_luminescence_Link_2015,Bouhelier_PL_2019,Baumberg-Aizpurua_Nat_comm_2024_inelastic,Harutyunyan_2024}, trimers~\cite{Joel_Yang_ACS_Photonics_2016} etc., or to other plasmonic materials, including low electron density Drude materials such as transparent conducting oxides~\cite{Kinsey-Nat-Rev-Mater-2019,Un-Sarkar-Sivan-LEDD-II}. We also emphasize that the analysis in the current manuscript is limited to CW illumination, but could be extended also to pulsed illumination~\cite{Beversluis_PRB_2003,Lupton_transient_metal_PL,Lupton_transient_metal_PL_2017,Hecht_PRL-2019,Suemoto_PRB_2019}. In this scenario, which was studied more extensively in the past, the size-dependence of the temperature is different, and there is a dynamic transition between the non-thermal and thermal parts (see~\onlinecite{Sivan_tPL}), which entails further complexity. This class of experiments will be analyzed separately in future work. Ultimately, our work paves the way to simple optimization of the PL for practical purposes.

\bigskip

{\bf Acknowledgements}. I. K. and Y.S. were partially funded by Israel Science Foundation (ISF) grant no. 340/2020, a Lower-Saxony - Israel collaboration grant no. 76251-99-7/20 (ZN 3637) as well as a grant from the Bronitzki family.
I.W.U. was funded by the Guangdong Natural Science Foundation (Grants No. 2024A1515011457).

\begin{suppinfo}
The Supporting Information is available free of charge.

Absorption cross-section of spheres, absorption characteristics across spectrum, thermal and non-thermal contribution to PL of sphere, PL of sphere excited off-resonance, PL calculated for higher order mode responses, the thermo-optic effect on absorption, PL calculation for films.

\end{suppinfo}

\providecommand{\latin}[1]{#1}
\makeatletter
\providecommand{\doi}
  {\begingroup\let\do\@makeother\dospecials
  \catcode`\{=1 \catcode`\}=2 \doi@aux}
\providecommand{\doi@aux}[1]{\endgroup\texttt{#1}}
\makeatother
\providecommand*\mcitethebibliography{\thebibliography}
\csname @ifundefined\endcsname{endmcitethebibliography}
  {\let\endmcitethebibliography\endthebibliography}{}

\end{document}